\documentclass[a4paper,11pt]{article}
\pdfoutput=1 

\usepackage{jheppub} 
\usepackage[utf8]{inputenc}
\usepackage{setspace}

\definecolor{refkey}{gray}{0.45}
\definecolor{labelkey}{RGB}{155,48,48}

\usepackage{physics}
\usepackage[obeyFinal]{todonotes}

\def\beq{\begin{eqnarray}}\def\eeq{\end{eqnarray}}
\def\be{\begin{equation}}\def\ee{\end{equation}}

\def\mes[#1]{d^{3}{#1}}

\def\del{\partial}

\newcommand{\half}{\frac{1}{2}}

\def\del{\partial}

\def\order{\ensuremath{\mathcal{O}}}

\reversemarginpar

 \author[a]{Upamanyu Moitra,}
  \author[a]{Sunil Kumar Sake,}
 \author[a]{Sandip P. Trivedi,}
 \author[a]{V. Vishal}
 \affiliation[a]{\it Department of Theoretical Physics,
 Tata Institute of Fundamental Research,\\  Colaba, Mumbai, India, 400005\\}

\emailAdd{upamanyu@theory.tifr.res.in}
\emailAdd{sunil.sake@tifr.res.in}
\emailAdd{sandip@theory.tifr.res.in}
\emailAdd{vishal@theory.tifr.res.in}

\vspace{1cm}

\abstract{We analyse the Jackiw-Teitelboim model of 2D gravity coupled to $N$ massless free scalar fields in the semi-classical limit. Two systems are studied which essentially differ in the boundary conditions that are imposed. We find that  the thermodynamics has interesting differences. 
 We also analyse the response to additional infalling matter which satisfies the null energy condition.   The second law is shown to be valid in both systems for  the generalised entropy which  takes into account the entanglement across 
the  event horizon due to the matter fields. 
Similarly we  find that the generalised entropy increases along  future Q-screens in both systems.  
}

\title{Jackiw-Teitelboim Model Coupled to Conformal Matter in the Semi-Classical Limit}

\preprint{\parbox{3cm}{TIFR/TH/19-27}}

\begin{document}
\maketitle
\flushbottom
\section{Introduction}
\label{intro}
The Jackiw-Teitelboim (JT) model \cite{JACKIW1985343,Teitelboim:1983ux} of 2D gravity -- see, for example, \cite{Muta:1992xw,Lemos:1993qn,Lemos:1996bq} for some early important work on the model -- has attracted wide attention lately \cite{alm, Jensen:2016pah, Engelsoy:2016xyb, Maldacena:2016upp, Cvetic:2016eiv, Maldacena:2017axo, Dubovsky:2017cnj, Taylor:2017dly, Kourkoulou:2017zaj, Grumiller:2017qao, Gonzalez:2018enk, Gaikwad:2018dfc, nayak, Kolekar:2018sba, Maldacena:2018lmt, Harlow:2018tqv, Li:2018omr, Bena:2018bbd, Larsen:2018iou, Goel:2018ubv, Lin:2018xkj, Castro:2018ffi, Callebaut:2018nlq, Kitaev:2018wpr, Moitra:2018jqs, Yang:2018gdb, Brown:2018bms, Larsen:2018cts,  Blommaert:2018iqz, Alishahiha:2018swh, Dhar:2018pii, Kolekar:2018chf, Goto:2018iay, Sachdev:2019bjn, Blommaert:2019hjr, Saad:2019lba, Mertens:2019bvy, Maldacena:2019cbz, Mertens:2019tcm, Iliesiu:2019xuh, Lin:2019qwu, Cotler:2019nbi, Moitra:2019bub, Castro:2019crn, Stanford:2019vob, Gross:2019ach, Hong:2019tsx, Sarosi:2017ykf, joshi2019time}. 
This model captures the behaviour of a class of statistical mechanics models called the SYK models \cite{Sachdev:1992fk, Kitaev-talks:2015, Polchinski:2016xgd, Maldacena:2016hyu, Jevicki:2016bwu, Danshita:2016xbo, Bagrets:2016cdf, Jevicki:2016ito, Gu:2016oyy, Gross:2016kjj, Berkooz:2016cvq, Garcia-Garcia:2016mno, Fu:2016vas, Witten:2016iux, Gurau:2016lzk, Klebanov:2016xxf, Davison:2016ngz, Peng:2016mxj, Krishnan:2016bvg, Turiaci:2017zwd, Li:2017hdt, Gurau:2017xhf, Mandal:2017thl, Bonzom:2017pqs, Gross:2017hcz,  Stanford:2017thb, Krishnan:2017ztz, Das:2017pif, Narayan:2017qtw, Chaudhuri:2017vrv, Murugan:2017eto, Krishnan:2017txw, Gross:2017vhb, Garcia-Garcia:2017bkg, Anninos:2017cnw, Giombi:2017dtl, Sonner:2017hxc, Bulycheva:2017ilt, Choudhury:2017tax, Gross:2017aos, Kitaev:2017awl, Das:2017hrt, Das:2017wae, Narayan:2017hvh, Haehl:2017pak, Garcia-Garcia:2018pwt, Krishnan:2018hhu, Roberts:2018mnp, Benedetti:2018goh, Klebanov:2018nfp, Gharibyan:2018jrp, Saad:2018bqo, Gubser:2018yec, Blommaert:2018oro, Chang:2018sve, Gur-Ari:2018okm, Liu:2018jhs, Giombi:2018qgp, Pakrouski:2018jcc, Blake:2018leo, Bhattacharya:2018fkq, Murugan:2018fdj, Kim:2019upg, Sun:2019mms, Guo:2019csw, Nayak:2019khe, Sun:2019yqp, Klebanov:2019jup, Rosenhaus:2018dtp, Chen_2019}   which have similarities with the behaviour of near-extremal black holes. 
In fact, it has been shown that the low-temperature and low-frequency behaviour of a wide class of near-extremal black holes, including rotating ones, is well approximated in a precise way by the JT model \cite{Maldacena:2016upp,nayak,Moitra:2018jqs,Moitra:2019bub,Sachdev:2019bjn}.

In this paper we consider  the behaviour of 2D theories obtained by coupling the JT model to $N$ extra massless scalar fields, $\psi_i,~ i=1, \cdots N$. We work in the semi-classical limit obtained by taking $N\rightarrow \infty$ and the 2D Newton's constant $G \rightarrow 0$,  keeping $N G$ fixed. 
The quantum  effects of matter are included in this approximation, while the gravity-dilaton  sector behaves classically. 

We examine two models here. 

In the first model the effect of the $N$ scalars is replaced by  one scalar 
called $\chi$ below, which is non-minimally coupled, with an action 
\be
\label{actchi}
I_\chi= -{N \over 24 \pi}  \int \sqrt{-g} \,\text{d}^2x \,\big[(\partial \chi)^2 + R \chi\big],
\ee
(and an appropriate boundary term, see eq.\eqref{nonimpact}). 
Due to the non-minimal coupling  this single field reproduces the conformal anomaly of the $\psi_i$ fields originally present.  The  prefactor 
in eq.(\ref{actchi}) scaling like $N$ also means that the field   $\chi$ is classical   in the large $N$ limit.  The  dynamics of the full system is then obtained by coupling this classical field to the classical gravity-dilaton system.  The spacetime we consider 
has a boundary where the dilaton takes a fixed value; we also impose vanishing boundary conditions on $\chi$ at this boundary.

In the second model we work directly with the $N$ scalar fields $\psi_i,~ i=1,\cdots N$ which are minimally coupled with an  action
\be
\label{acts}
I_\psi=\frac{1}{2}\int \sqrt{-g}\,\text{d}^2x\, \sum_{i=1}^N ( \partial \psi_i)^2,
\ee
and impose Dirichlet boundary conditions on these scalars.

We find that the two models have interesting differences, essentially due to the different boundary conditions which are imposed on them.
In both cases, infalling matter  results in the formation of a black hole which  evaporates and eventually settles down to thermal equilibrium.
In this equilibrium state, matter is radiated by the black hole, bounces off the boundary and eventually falls back into the black hole, with the rate 
of Hawking evaporation equalling that of the infalling matter energy-momentum. 

Quantum corrections result in corrections to the free energy  and mass of the black hole as a function of temperature; these are different for  the two systems. 
The first law of thermodynamics is obeyed in the presence of these corrections once the entropy is replaced by  the generalised entropy which  also includes a contribution coming from entanglement across the horizon. 

For time-dependent situations, we find that once the  external sources are turned off,  the system relaxes to thermal equilibrium. In the $\chi$ system 
this relaxation is due to a quasi-normal mode which decays exponentially with an exponent which depends on the temperature $T$ and $GN$. In the $\psi$ system the system relaxes instantaneously. 

Finally, we find that in both cases the Second Law of Thermodynamics is obeyed for the generalised entropy in the presence of additional classical matter which satisfies the null energy condition. We also analyse the behaviour of future Q-screens, which are an analogue of the locus swept out by the apparent horizon and find that the generalised entropy increases along the  future Q-screen in both models. See, for instance,  \cite{Wall:2009wm, Wall:2009wi, C:2013uza, Wall:2011hj, Wall:2011kb,  Bousso:2015qqa, boussonewarea,  bousso1506qfc,  Bousso:2015wca, boussogencosmo, Engelhardt:2017aux, Wall:2018ydq, Bousso:2018fou}, and the references therein for some of the relevant literature.



This paper is structured as follows. In section 2 we review the classical behaviour of the JT model, including the black hole solution and the response to general infalling matter. The behaviour of the $\chi$  and $\psi$ systems is considered in section 3 and 4 respectively. The results of entropy monotonicity along Q-screen are discussed in section 5.  Finally we end with conclusions in section 6. Appendices \ref{appcoord}-\ref{cldlt} contain important additional details. 

\section{Basic Setup}
\label{basic}
The Jackiw-Teitelboim (JT) model  consists of 2D gravity coupled to a scalar, $\phi$, called the dilaton,
with an action, 
\begin{align}
I_{JT}=\frac{1}{16\pi G}\pqty{\int d^2 x\,\sqrt{-g}\,\phi (R-\Lambda_2)+2\int_{bdy}\sqrt{-\gamma}\phi K-\frac{2}{L_2}\int_{bdy}\sqrt{-\gamma}\phi }\label{jtact}.
\end{align}
Here $\Lambda_2$ is the 2D cosmological constant given by $\Lambda_2=-\frac{2}{L_2^2}$. The last term in the eq.\eqref{jtact} is a counter-term which is required to remove  divergences  that arise while computing  the on-shell action and related quantities. We shall set the $AdS_2$ length $L_{2}=1$ in the calculations to follow. 
Let us work in the conformal gauge in which the metric takes the form,
\begin{align}
\text{d}s^2 =-e^{2\omega(x^{+},x^{-})}\,\text{d}x^{+}\text{d}x^{-}\label{conmet}.
\end{align}
The equation of motion by varying the dilaton in the above action is
\begin{equation}
4 \del_{+}\del_{-}\omega+e^{2\omega}=0\label{dileq}.
\end{equation}
It is easy to see that the equation above has a solution in which the spacetime is  $AdS_2$, with the metric in Poincar\'e coordinates $(t,z)$ (related to the $x^\pm$ coordinates as $x^\pm = t \pm z$)  given by
\be
\text{d}s^2={1\over z^2}(-\text{d}t^2+\text{d}z^2)=  -\frac{4}{(x^{+}-x^{-})^2} \text{d}x^{+}\text{d}x^{-}\label{poinmet}.
\ee
Varying the metric we obtain, 
\begin{equation}
\frac{1}{8\pi G}(\nabla_{\mu}\nabla_{\nu}\phi-g_{\mu\nu}\nabla^{2}\phi+ g_{\mu\nu}\phi)=0.\label{purejtphi}
\end{equation}
This admits, as one of the solutions, a linearly varying dilaton
\be
\label{linvardil}
\phi={1\over 2{\cal J} z},
\ee
where ${\cal J}$ is an energy scale. The linear variation  of the dilaton breaks the $SL(2,R)$ isometries of $AdS_2$ to $U(1)$ \cite{Maldacena:2016upp} and ${\cal J}$ characterises the scale of this breaking. 

The  spacetime has a boundary where the dilaton takes a fixed value
\be
\label{bdrydil}
\phi=\phi_B.
\ee

\subsection{Vacuum Solutions}
\label{vacuumsols}
The general solution for the dilaton satisfying eq.\eqref{purejtphi} is 
\be
\label{solid}
\phi={a+b(x^+ + x^-)+ cx^+x^-
\over \mathcal{J}( x^+-x^-)},
\ee
where $a,b,c$ are arbitrary constants. It is easy to show that when  the parameters, $a,b,c$  meet the following two conditions,
\begin{align}
   \mu  & \equiv b^2-ac > 0 \label{muabc}, \\
 \phi_B &>  {\sqrt{\mu}\over \mathcal{J}}  \label{bgh},
 \end{align}
   the solution describes a black hole. The steps detailing the calculation of mass are shown in appendix \ref{appclass} with  the mass given by 
   \be
   \label{valmass}
   M=\frac{\mu}{16\pi G\,\mathcal{J}},
   \ee
   Doing an appropriate $SL(2,R)$ transformation brings eq.\eqref{solid} to the form eq.\eqref{dil1}. Further coordinate transformations shown in appendix \ref{appcoord} can be done to bring the metric and the dilaton to the form eq.\eqref{bhmet3} and eq.\eqref{dilr} respectively. It then follows immediately that the temperature of the black hole is given by 
   \be
   \label{valT}
   T=\frac{\sqrt{\mu}}{2\pi}.
   \ee
   As discussed in \cite{Engelsoy:2016xyb,Maldacena:2016upp} the mass can be expressed in terms of the Schwarzian action.
   Let the proper time along the boundary be\footnote{More correctly, in Fefferman-Graham coordinates the metric near the boundary takes the form, $\mathrm{d}s^2= -({1+ \mathcal{O} ({\hat z})^2\over {\hat z}^2}) \mathrm{d} {\hat t}^2  + {\mathrm{d} \hat {z}^2\over {\hat z}^2}$, and 
   $\phi={1\over {\cal J} {\hat z} } $.} ${\hat t}$, the boundary can be described by the function $t({\hat t})$ and the mass is given by,
   \be
   \label{masssch1}
   M=-{1\over 8 \pi G \mathcal{J} }\text{Sch}(t, \hat t),
   \ee
   where 
   \begin{equation}
   \text{Sch}(t, \hat t ) \equiv {t'''( \hat t)\over t'( \hat t)}-{3\over 2}\pqty{t''( \hat t )\over t'( \hat t )}^2. \label{schderiv}
   \end{equation}
  When the JT model arises from higher dimensional theories, $\phi$ is related to the area of the transverse sphere along the additional dimensions and its value at the horizon is  proportional to  the increase in  area when the black hole is made non-extremal. This motivates the definition of the black hole entropy in the JT model to be 
   \be
   \label{defentbh}
   S_{BH}= {\phi\over 4 G}\bigg\vert_{h}.
   \ee
  The horizons lies along the loci where $(\nabla \phi)^2=0$. From eq.(\ref{solid})  it is easy to see that the future and past event horizons  which correspond to the conditions 
  $\partial_-\phi=0$ and $\partial_+\phi=0$ respectively lie at 
  \be
  \label{locha}
 x_h^+={-b-\sqrt{\mu}\over c},
  \ee
  and 
  \be
  \label{lochb}
 x_h^-={-b+\sqrt{\mu}\over c}.
  \ee
  
This leads to  entropy, eq.(\ref{defentbh}),
\be
\label{entbha}
S_{BH}= \frac{\sqrt{\mu}}{4G\,\mathcal{J}}.
\ee
From eq.(\ref{valmass}) ,  eq.(\ref{valT}) and eq.(\ref{entbha}) it is easy to see that the first law of thermodynamics
\be
\label{flaw}
T\text{d}S_{BH} = \text{d}M ,
\ee
 is obeyed by these black holes.

It is worth noting that our calculation for the mass leading to eq.(\ref{valmass}) is carried out using the holographic renormalisation method and as explained in appendix \ref{adm} it  is valid only for small temperatures,
\be
\label{condo}
{T\over \mathcal{J}} \ll \phi_B.
\ee
In this limit we are dealing with ``small" black holes whose horizon radius is deep inside the boundary and  the value of the dilaton at the event horizon $\phi_h$ meets the condition,
\be
\label{conddilh}
\phi_h\ll \phi_B.
\ee
However the agreement with the first law shows that the above expression for the mass eq.(\ref{valmass}) should be valid for bigger black holes, at higher temperatures,  as well. 

\subsection{Infalling Matter}
\label{infall}
Let us next couple the JT model to classical matter. 
In the presence of matter the metric equations of motion become, 
\begin{equation}
- \frac{1}{8\pi G}(\nabla_{\mu}\nabla_{\nu}\phi-g_{\mu\nu}\nabla^{2}\phi+ g_{\mu\nu}\phi)=T^{m}_{\mu\nu}\label{geneqs},
\end{equation}
with $T^{m}_{\mu\nu}$ being the matter stress tensor.

In the conformal gauge, (\ref{conmet}), the components of
 eq.\eqref{geneqs} become
\begin{align}
-e^{2\omega}\del_{\pm}\pqty{e^{-2\omega}\del_{\pm}\phi}&=8 \pi G T^m_{\pm\pm}\label{ppmm},\\
2\del_{+}\del_{-}\phi+e^{2\omega}\,\phi&=16\pi G T^m_{+-}\label{pm}.
\end{align}
We consider the special case of conformally invariant matter, with $T^{m}_{+-}=0$,  which does not  couple to the dilaton but only to the metric.
Varying the dilaton in eq.\eqref{jtact}, we get once again, eq.(\ref{dileq}),
which leads to the spacetime being  $AdS_2$ as before. Let us also  take the matter stress tensor to be  purely infalling, i.e., $T^m_{++}=0$. In summary,
\be
\label{condstress}
T_{+-}^m=0 = {T_{++}^m}.
\ee
Also we take the $T_{--}^m$ component to satisfy 
\be
\label{energycond}
T_{--}^m>0.
\ee
This is true if the matter satisfies the null energy condition (NEC)
\begin{equation}
T^m_{ab} k^a k^b>0\label{neca},
\end{equation} 
with $k^a$ being the future-directed tangent vector for  any null geodesic,
\be
\label{tannull}
k^a = {\text{d}x^a \over \text{d}\lambda}, 
\ee
where $\lambda$ is an affine parameter along the geodesic.
The `$++$' equation of (\ref{ppmm})   leads to, 
\begin{equation}
\phi= d(x^{-})+\frac{h(x^{-})}{x^{+}-x^{-}}.\label{dhphi}
\end{equation}
Then eq.(\ref{pm}) allows $d(x^-)$ to be determined in terms of $h(x^-)$ leading to 
\begin{equation}
\phi= \frac{1}{2}\partial_{-}{h}(x^{-})+\frac{h(x^{-})}{x^{+}-x^{-}}\label{gendil}.
\end{equation}
The remaining `$--$'  component of (\ref{ppmm}) then gives, 
\be
\label{mmcim}
-{1\over 2} h'''= 8 \pi G T_{--}^m.
\ee
Let us note that eq.(\ref{mmcim}) follows from the boundary action 
\be
\label{bait}
I= -{1\over  {8 \pi G \cal J}} \int \mathrm{d} {\hat t} \, \mathrm{Sch} (t, \hat t ) + I^m,
\ee
where ${\hat t}$ is the boundary time and the boundary stress tensor $T^m_{\hat t \hat t}$ is obtained from the matter action $I^m$, when expressed as a functional of $t(\hat{t})$ on the boundary, as
\be 
\label{tttmat}
T^m_{\hat t \hat t} =  t' \frac{\delta I^m }{ \delta t(\hat t) } .
\ee This follows from eq.(\ref{relthatt}) as  is discussed further  in appendix \ref{appclass}. 

Note that for $T_{--}^m=0$, eq.(\ref{mmcim}) has a general solution 
\be
\label{genvac}
h_0={1\over \mathcal{J}}{(a+2bx^-+ c(x^-)^2)},
\ee
for some constants $a, b, c$. It is easy to see that this, in conjunction with eq.\eqref{gendil},  agrees with eq.(\ref{solid}) above. 

Consider a situation where  we start for $x^-<0$ with a solution of the form eq.(\ref{genvac}) corresponding to  a black hole and then allow  matter  to fall in. It follows from 
eq.(\ref{mmcim}) that  
the resulting solution is given by 
\be
\label{gensolaa}
h=h_0-{16 \pi G}\int_{x^-=0}\int_{x^-=0}\int_{x^-=0} T_{--}^m .
\ee

The initial  black hole  mass is given in eq.(\ref{valmass}). The additional matter leads to a further increase in mass given by \footnote{Noting that $x^{\pm}=t\pm z$ and using eq.\eqref{condstress}, \eqref{relthatt}, we see that
\begin{equation}
\label{mpropt}
\Delta M={\, \mathcal{J}\over 2} \int \text{d}x^- h\, T_{--}^m={\, \mathcal{J}\over 2} \int \text{d}t\,\, h(t)\, T_{tt}^m=\int \text{d}\hat{t}\,\, T^m_{\hat{t}\hat{t}}\nonumber.
\end{equation}
}
\be
\label{incmass}
\Delta M={\, \mathcal{J}\over 2} \int \text{d}x^- h\, T_{--}^m.
\ee
The total mass is given in terms of $h$ (see  appendix \ref{appclass})   by 
\begin{equation}
  M=\frac{\mathcal{J}}{64\pi G}\left(h'^{2}-2 h h''\right).\label{admclass}
  \end{equation}
In the above expression, $h$ is evaluated at $z\rightarrow 0$ and prime indicates a derivative with respect to the Poincar\'e time $t$, see appendix \ref{adm} for more details. The calculation in appendix \ref{adm} is justified for black holes which are small, i.e. for which the value of the dilaton at the horizon meets the condition, eq.(\ref{conddilh}). 

As is discussed in appendix \ref{appclass} in the more general time dependent case also, $M$ can be expressed in terms of the Schwarzian derivative as given in 
eq.\eqref{masssch}. This is to be expected, since the Schwarzian term is the simplest one consistent with  $SL(2,R)$ symmetry.

From eq.(\ref{bdrydil})  and eq.(\ref{gendil}) we see that the boundary of spacetime can be expressed as a function of $x^-$, 
\be
    x^+ = x^- + \frac{2h(x^-)}{2\phi_B-h'(x^-)}\label{xpxmrel}.
\end{equation}

We see from eq.(\ref{gensolaa})  and eq.\eqref{energycond} that $h$ becomes smaller as $x^-$ increases. Generically $h$ cannot go to 
zero in finite proper time as measured on the boundary. This follows from the fact that boundary proper time goes as 
${\hat t} \sim \int {\text{d}x^{ -}\over h}$, and so near a first order zero, $x_0$, of $h$, 
\begin{equation}
{\hat t} \sim - \ln(x_0-x^-)    \label{bdytime}.
\end{equation}
It is therefore enough to only consider the evolution till $h$ hits a zero, since matter cannot fall in thereafter from the boundary.

\subsection{Second Law }
\label{secondclass}
Here we verify that for  infalling matter of the type considered in the previous subsection, the second law is valid as long as the energy condition 
eq.(\ref{energycond}) is satisfied. More specifically, we show below that the value of the dilaton increases monotonically along the event horizon. We had mentioned above that the horizon value of the dilaton plays the role of the horizon area and its monotonic increase is therefore the analogue of the area increase theorem in this system. 

For simplicity we consider situations where the matter falls in for some duration (i.e. some interval in $x^-$) and then stops. 
The solution  after the matter stops falling is  then  of the form, eq.(\ref{solid}), eq.(\ref{genvac}), and has an event horizon $(eh)$ located at  eq.(\ref{locha}).
We can choose  an affine parameter $\lambda$ along the event horizon which meets  the condition 
\be
\label{affine}
{\text{d}x^-\over \text{d}\lambda}= (x_h^+-x^-)^2.
\ee
It then follows that the first derivative,
\be
\label{fderv}
{\text{d}\phi \over \text{d}\lambda}= (x_h^+-x^-)^2 {\text{d}\phi \over \text{d}x^-}\bigg\vert_{eh}.
\ee
It is then easy to see  that the second derivative satisfies the condition
\be
\label{sderv}
{\text{d}^2\phi\over \text{d}\lambda^2}=(x^+_h-x^-)^4 \nabla_- \nabla_- \phi\big\vert_{eh} = -8 \pi G (x^+_h-x^-)^4 T_{--}^m< 0,
\ee
where the second equality is due to the `$--$' equation of motion  (\ref{ppmm}) and  the last inequality follows from eq.(\ref{energycond}). 

Once  the matter stops falling in, at late times,  
\be
\label{condorp}
{\text{d}\phi \over \text{d}x^-}\bigg\vert_{eh}=0,
\ee
 since the event horizon is at $x^+=x^+_h$ {and the value of the dilaton at the event horizon is independent of $x^-$}.  
 From the inequality, eq.(\ref{sderv}), it then follows that ${\text{d}\phi \over \text{d}\lambda}>0$ along the event horizon, and thus $\phi$ monotonically increases.
  
 \subsubsection{The Apparent Horizon}
 \label{clasappar}
 It is interesting to note that in the classical system being analysed in this section the value of the dilaton at the apparent horizon $(ah)$ also increases monotonically.
 This will however turn out not to be necessarily true once we include the quantum effects of matter. 
 At the apparent horizon,
 \be
 \label{condapph}
 \partial_-\phi |_{ah}=0.
 \ee
 From eq.\eqref{gendil},  this leads to the condition
 \begin{equation}
 x^{+} |_{ah} = x^{-}-\frac{h'\pm\sqrt{h'^{2}-2h h''}}{h''}.\label{xpxmrela}
   \end{equation}
 Requiring the value of the dilaton at the apparent horizon to be non-negative selects the upper sign for the discriminant, resulting in    
   \begin{align}
   x^{+}\,|_{ah}&=x^{-}-\frac{h'+\sqrt{h'^{2}-2h h''}}{h''},\label{xphor}\\
  \phi\,\vert_{ah}&=\frac{1}{2}\sqrt{h'^{2}-2h h''}\label{phihor}.
  \end{align}
  
  We can parametrise the trajectory of the apparent horizon by the coordinate $x^-$ itself. 
  We have, 
    \begin{equation}
  \frac{\text{d}\phi}{\text{d}x^-}\bigg\vert_{ah}=\pqty{\frac{\partial \phi}{\partial x^{-}}+\frac{\partial\phi}{\partial x^{+}}\frac{\partial x^{+}}{\partial x^{-}}}\bigg\vert_{ah}=\left(\frac{-h}{ (x^{+}-x^{-})^{2}}\right)\frac{\partial x^{+}}{\partial x^{-}}\bigg\vert_{ah}\label{phiratehor},
  \end{equation}
  where we used eq.\eqref{condapph} and eq.\eqref{gendil}.
  But we also have $\frac{\text{d}(\del_{-}\phi)}{\text{d}x^-}=0$ along the apparent horizon. This gives
  \begin{align}
  \del_{-}x^{+}|_{ah}=-\frac{\del^{2}_{-}\phi}{\del_{+}\del_-\phi}\bigg\vert_{ah}=-\frac{16\pi G T_{--}^m}{e^{2\omega}\phi}\bigg\vert_{ah},\label{delxp}
  \end{align}
  where  we used the equations of motion eqs.\eqref{ppmm},\eqref{pm}, along with eq.\eqref{condstress}  and eq.(\ref{condapph}). Therefore using eq.\eqref{delxp} and the Poincar\'e metric eq.\eqref{poinmet}, eq.\eqref{phiratehor} becomes
  \begin{equation}
  \frac{\text{d}\phi}{\text{d}x^-}\bigg\vert_{ah}=4\pi G\,\frac{ h\,T_{--}^m }{\phi}\bigg\vert_{ah}\label{phirate}.
  \end{equation}
 Now we note that the RHS in this equation is positive, since the energy condition, eq.(\ref{energycond}) is met and $\phi$ and $h$ are positive. 
 It therefore follows that the dilaton increases with increasing $x^-$, which is along the direction of increasing time.  We also mention that while $h$ 
 decreases with increasing $x^-$, eq.(\ref{gensolaa}), it cannot turn negative in a finite duration of boundary time, see comments above after eq.(\ref{xpxmrel}).  Finally, we also note from eq.(\ref{delxp}) that in the presence of infalling matter the apparent horizon is 
 space-like.

\section{Semi-Classical Analysis: The $\chi$ System}
\label{quant}

We now turn to studying the effects of  the  quantum backreaction due to the matter fields.  We will  consider a simple system consisting of 
$N$ massless scalar fields which are  minimally coupled with an the action \eqref{acts}.
As was discussed in the introduction, we will work in the semiclassical limit obtained by taking $G \rightarrow 0$ and $N\rightarrow \infty$, keeping 
their product $GN$ fixed. In this limit,  quantum effects of the matter fields need to be included while  those  of  the gravity-dilaton fields can be neglected, 
since they are suppressed by $G$ and not enhanced by a  factor of $N$. For later convenience, we find it useful to define the parameter
\begin{equation}
\zeta=\frac{4 G N}{3}\label{zeta}.
\end{equation}

Once quantum effects are included it is well known that the free scalar theory has a conformal anomaly. As a result, for the matter system above we get 
\be
\label{tracing}
T^{\mu}_\mu= \frac{N}{24\pi} R,
\ee
where $R$ is the Ricci scalar. 

The same value for $T^\mu_\mu$  can be obtained in a classical theory of a single scalar $\chi$ which is non-minimally coupled with an action:
\begin{align}
I_{\chi}&=-\frac{N}{24\pi}\pqty{\int \text{d}^2 x\sqrt{-g}\pqty{\del_{\mu}\chi\del^{\mu}\chi+\chi R}+2\int_{bdy}\sqrt{-\gamma}\chi K}\label{nonimpact}.
\end{align}
Note that the action has a prefactor which goes like $N$. It is easy to see that the resulting stress energy tensor gives $T^\mu_\mu$ is in agreement with eq.(\ref{tracing}). 

One way to include the quantum effects of the $\psi_i$ fields  is therefore to work with the classical $\chi$ theory, eq.(\ref{nonimpact}), and couple it to the JT model. 
This system was analysed in \cite{alm} and we will study it first in this section. Following this in the next section, we will return to studying the system of the $\psi_i$ fields directly. 

 To completely specify the dynamics of this system we also need to specify the boundary condition satisfied by $\chi$.   We will take $\chi$ to satisfy Dirichlet boundary conditions (specifically, $\chi_B =0$) at the boundary of spacetime where $\phi=\phi_B$.
 
\subsection{Thermodynamics}
\label{chitherm}
To begin,  we briefly review the thermodynamics of the JT model coupled to the matter theory with action eq.(\ref{nonimpact}). This was analysed in detail in \cite{alm}.

Since the $\chi$ field does not couple to the dilaton, varying the dilaton again gives eq.\eqref{dileq} and so the geometry still remains $AdS_2$. The  black hole solution is  conveniently described by writing the $AdS_2$ metric in Schwarzschild coordinates:  
\be
\label{metic}
\text{d}s^2=-(r^2-\mu) \,\text{d}t_s^2 + {\text{d}r^2\over (r^2-\mu)}.
\ee
The parameter $\mu$  determines the mass etc. of the black hole. 
The temperature is given by the expression (\ref{valT})
The metric eq.(\ref{metic}) is independent of $t_s$ and therefore a shift of  the $t_s$ coordinate is manifestly a symmetry. 

To proceed it is convenient to go to conformal coordinates, 
\be
\label{defxpmther}
x_s^\pm\equiv t_s\pm r_*,
\ee
 where 
\be
\label{dears}
r_*=-\int {\text{d}r\over r^2-\mu} = -{1\over  2\sqrt{\mu}}\ln(r-\sqrt{\mu}\over r+ \sqrt{\mu}).
\ee
In these coordinates the conformal factor, $\omega$, which appears in the metric eq.\eqref{conmet}
is given by 
\be
\label{valomega}
\omega=\frac{1}{2}\,\ln(r^2-\mu).
\ee
The equation of motion for $\chi$,  from the action eq.\eqref{nonimpact} is,
\be
\del_+\del_-(\chi+\omega)=0\label{chieom},
\ee
 which gives, 
\be
\label{solchi}
\chi=-\omega+f_+(x^+_s) + f_-(x^-_s) + c_0,
\ee
for some arbitrary functions $f_{+}$ and $f_{-}$. Here $c_0$ is a constant.  Requiring $\chi$ to be independent of $t_s$ fixes
\begin{align}
\label{fpm}
f_+(x^+_s) & = p\, x_s^+, \\
f_-(x^-_s) & = -p\, x_s^-\label{fmm},
\end{align}
for some constant $p$.
The components of the stress tensor for the $\chi$ field are,
\begin{align}
T_{+-}^{\chi}&=\frac{N}{12\pi} \partial_{+}\partial_{-}\chi\label{Tpmchi},\\
T_{\pm\pm}^{\chi}&=\frac{N}{12\pi}\left(-\partial^{2}_{\pm}\chi+\partial_{\pm}\chi\partial_{\pm}\chi+2\del_{\pm}\chi\del_{\pm}\omega\right)\label{Tppchi}.
\end{align}
Using the solution eq.\eqref{solchi}, eq.\eqref{fpm} and eq.\eqref{fmm}, the stress tensor components become
\begin{align}
T_{+-}^{\chi}&=-\frac{N}{12\pi} \partial_{+}\partial_{-}\omega\label{Tpm},\\
T_{\pm\pm}^{\chi}&=\frac{N}{12\pi}\left(\partial^{2}_{\pm}\omega-\partial_{\pm}\omega\partial_{\pm}\omega\right)+ \frac{N}{12\pi}p^2\label{Tpp}.
\end{align}

 From the equations of motion for the `$+ -$' component of the metric, eq.\eqref{pm} with $T_{+-}^m$ replaced by eq.\eqref{Tpm},  we learn that 
 \be
 \label{behdil}
 \phi={ r \over {\cal J}}+\frac{\zeta}{4},
 \ee
 where we have used the definition (\ref{zeta}). Here we have imposed that $\phi$ is independent of $t_s$ and regular at the horizon\footnote{For $\phi$  which is independent of  $t_s$ eq.(\ref{pm}) reduces to an ordinary second order equation in $r$ which has only one solution regular at the horizon.}, which is given by
 \begin{equation}
 r=\sqrt{\mu}.\label{horizon}
 \end{equation} 
 $\mathcal{J}$ is the energy scale which breaks the scaling symmetry and is the same that appears in eq.\eqref{linvardil}.
  The `$++$' and `$--$' components of the equations of motion (\ref{ppmm}) with $T_{\pm\pm}^m$ replaced by $T_{\pm\pm}^{\chi}$, eq.\eqref{Tpp}, then determines
 \be
 \label{valbeta}
 p = -{\sqrt{\mu}\over {2}}.
 \ee
 Note that with this value of $p$, the field $\chi$ is given by  
 \be
 \label{chill}
 \chi=-\ln(r+\sqrt{\mu}) + c_0,
 \ee
 and is non-singular at the horizon \eqref{horizon}.
 Finally demanding that $\chi$ vanish at the boundary $\phi=\phi_B$ gives, 
 \be
 \label{valco}
 c_0=\ln({\tilde \phi}_B\mathcal{J}+\sqrt{\mu}).
 \ee
 Here ${\tilde \phi}_B$ is given by 
 \be
 \label{valtphib}
 {\tilde \phi}_B= \phi_B -{\zeta\over 4}.
 \ee
 We will be  working in the limit where eq.(\ref{condo}) is met. In this limit 
 \be
 \label{limo}
 c_0\simeq \ln(\tilde{\phi}_B\mathcal{J}),
 \ee
 so that 
 \be
 \label{chilla}
 \chi=-\ln(r+ \sqrt{\mu}) + \ln(\tilde{\phi}_B\mathcal{J}).
 \ee

The ADM mass corresponding to the solution  can be calculated  as discussed in appendix \ref{adm}
and is given by 
\be
\label{masschi}
M= \frac{\mu}{16\pi G\, \mathcal{J}}+\frac{N\sqrt{\mu}}{12\pi}.
\ee
The computation of the ADM mass shows that  a counter-term is required to be added to the action eq.\eqref{nonimpact} to cancel divergences, 
\be
\label{counter term}
I^{\chi}_{ct} =-\frac{N}{24\pi} \int_{\del}\sqrt{-\gamma}.
\ee

We saw in the discussion of the previous section that the value of the dilaton at the horizon is analogous to the horizon area of higher dimensional gravity systems. 
Here we note that the $\phi$ and $\chi$ fields both couple to the curvature $R$ in the action, eq.\eqref{jtact} and \eqref{nonimpact}. This motivated a definition of the generalized 
entropy \cite{alm}
\be
\label{entropchi}
S^{\chi}_{gen}= \frac{1}{4G}\phi |_h -\frac{N}{6} \chi |_h,
\ee
where $\phi |_h$ and $\chi |_h$ refer to the horizon values of the two fields. 

Using eq.\eqref{behdil} and eq.\eqref{chilla}, and noting that event horizon is given by eq.\eqref{horizon}, we see that for the black hole solution under consideration,

\be
\label{entbchi}
S^{\chi}_{gen}= \frac{1}{4G}\pqty{{\sqrt{\mu}\over \, \mathcal{J}}+{\zeta\over 2}\pqty{\ln\sqrt{\mu}+\frac{1}{2}-\ln(\frac{\tilde{\phi}_B\mathcal{J}}{2})}}.
\ee

It is easy to see from eq.\eqref{entbchi}, \eqref{masschi} and \eqref{valT} that the system satisfies the first law of thermodynamics, $T\text{d}S^{\chi}_{gen}=\text{d}M$. Moreover, using eq.\eqref{masschi}, eq.\eqref{entbchi} and eq.\eqref{valT}, we see that the temperature dependent part of the quantity $\beta F$, where $F=M-T S^{\chi}_{gen}$ is the free energy and $\beta=1/T$, is given by 
\begin{equation}
\label{betaF}
\beta F=\beta M-S=-{\sqrt{\mu}\over 8 G \mathcal{J}}-{\zeta\over 8 G}\ln\sqrt{\mu}.
\end{equation}
Comparing this with the result obtained for the genus zero partition function obtained in \cite{Saad:2019lba}, we find that the value of $\beta F$ in eq.(122) in \cite{Saad:2019lba} corresponds to $N=9$ in eq.\eqref{betaF}, noting eq.\eqref{zeta}.

Let us end with one comment. As mentioned in section \ref{vacuumsols}, $\phi$ can be thought of as  being analogous  to  the area of the horizon in this 
model. The extra contribution due to the $\chi$ field in $S^{\chi}_{gen}$, eq.\eqref{entropchi}, can  be thought of as arising due to the entanglement of the $\psi_i$ 
matter fields across the horizon. The   quantum effects  of these fields have  been replaced   by the  classical $\chi$ field here and their entanglement
is replaced by the  value of $\chi$ at the horizon.   It is also worth noting that both
$(\nabla \phi)^2 $ and $(\nabla ({\phi\over 4 G}-{N \chi \over 6}))^2$ vanish at the horizon.

\subsection{Infalling Matter}
\label{infallchi}
Next we consider adding additional matter of the type considered in section \ref{infall}. The matter is conformal and taken to be classical. We analyse how the behaviour of the system changes due to the additional effects of the $\chi$ field. 

The matter  does not couple to the dilaton and its stress tensor $T^m_{\mu\nu}$ satisfies the conditions, eq.(\ref{condstress}) with $T^m_{--}$ being the only non-zero component. 

The dilaton equation of motion shows that the metric continues to be  $AdS_2$.
The stress tensor for the $\chi$ field in the conformal gauge eq.\eqref{conmet} is given in eq.\eqref{Tpmchi},\eqref{Tppchi}.
The equation of motion by varying $\chi$ is given in eq.\eqref{chieom} with the solution
\be
\label{solchia}
\chi= -\omega + f_+(x^+) + f_-(x^-)+ \ln(2 {\cal J} {\tilde \phi_B}),
\ee
where ${\tilde \phi}_B$ is defined in eq.(\ref{valtphib}), and the last term on RHS is added to simplify the following discussion.
Using eq.\eqref{solchia} in eqs.\eqref{Tpmchi}, \eqref{Tppchi}  we get 
\begin{align}
T_{+-}^{\chi}&=-\frac{N}{12\pi} \partial_{+}\partial_{-}\omega\label{Tpmanomaly},\\
T_{\pm\pm}^{\chi}&=\frac{N}{12\pi}\left(\partial^{2}_{\pm}\omega-\partial_{\pm}\omega\partial_{\pm}\omega\right)+\frac{N}{12\pi}\left(-\del^{2}_{\pm}f_{\pm} + (\del_{\pm}f_{\pm})^{2}\right)\label{Tppmm}.
\end{align}
Here after, in this section, we work in Poincar\'e coordinates, eq.\eqref{poinmet}.
In this case,  $\del^{2}_{\pm}\omega-(\del_{\pm}\omega)^{2}=0$, leading to
\begin{equation}
T^{\chi}_{\pm\pm}=\frac{N}{12\pi}\left(-\del^{2}_{\pm}f_{\pm} + (\del_{\pm}f_{\pm})^{2}\right)\label{tpmexp}.
\end{equation}

For simplicity we consider situations where we start with empty $AdS_2$  into which the additional matter begins to fall  at $x^-=0$.
Before that, for $x^-<0$, the $\chi$ field is  given by 
\be
\label{valchia}
\chi=- \omega + \ln(2 {\cal J} {\tilde \phi_B}),
\ee
with $f_+(x^+)=f_-(x^-)=0$ and satisfies the Dirichlet boundary condition.
Also, the dilaton is  given by, 
\be
\label{valdia}
\phi={1\over 2{\cal J} z} + {\zeta\over 4}.
\ee
It is easy to see that these solve the equations of motion. The term `Poincar\'e vacuum' will be used to refer to the initial configuration eq.\eqref{valchia} and eq.\eqref{valdia} in the discussion related to $\chi$ system in the rest of the paper.


Once the additional matter begins to fall in, $T^m_{--}$ no longer vanishes. However since $f_+(x^+)$ vanishes, $T^\chi_{++}=0$
and we learn from eq.(\ref{condstress})  that 
\be
\label{finalvaltpp}
T_{++}= T^m_{++}+T^\chi_{++}=0.
\ee
From the equations of motion for the `$+-$' (\ref{pm}), `$++$' eq.(\ref{ppmm}) components of the metric with $T_{\mu\nu}^m$ replaced by $T_{\mu\nu}^{\chi}+T_{\mu\nu}^{m}$ and using eq.\eqref{Tpmanomaly}, eq.\eqref{finalvaltpp},  we learn that the dilaton is determined in terms of one function of 
 $x^-$ coordinate, $h(x^-)$ by,
\begin{align}
	\phi= \frac{h'(x^-)}{2}+\frac{h(x^-)}{x^{+}-x^{-}}+{\zeta\over 4}\label{gendilq}.
	\end{align}
The value for $\phi$ given by eq.\eqref{valdia} corresponds to 
\be
\label{meh}
h={1\over {\cal J}}.
\ee
The remaining equation involving $T_{--}$ determines $h$ as
	\begin{equation}
	h'''=-16\pi G\,T^{\chi}_{--}-16\pi G\,T^{m}_{--}\label{hchi}.
	\end{equation}
 As was discussed above, we impose Dirichlet boundary condition for the field $\chi$ at the boundary $ \phi=\phi_B$. From eq.\eqref{gendilq}, it follows that the boundary trajectory is given by
\begin{equation}
z_B(x^-)=\frac{h}{2\tilde{\phi}_B-h'}\label{bdytraj},
\end{equation}
where ${\tilde \phi}_B$ is given in eq.(\ref{valtphib}).
As matter falls in, the form of $\chi$ is given by eq.\eqref{solchia} with $f_+=0$. By requiring that $\chi$ vanish at the boundary, we get 
\begin{align}
f_-=-\ln{\pqty{\frac{h}{2\tilde{\phi}_B-h'}}}-\ln\pqty{2\tilde{\phi}_B\mathcal{J}}\label{fminus},
\end{align}
leading to  
\begin{equation}
\chi=\ln{z}-\ln{\pqty{\frac{h}{2\tilde{\phi}_B-h'}}},\label{chisol2}
\end{equation}
where we used $\omega=-\ln z$ for the Poincar\'e metric, eq.\eqref{poinmet}.
Therefore, from eq.\eqref{tpmexp}, the $T^{\chi}_{--}$ component of the stress tensor becomes
\begin{align}
T_{--}^{\chi}=\frac{N}{12\pi}\frac{ \left(h h''' \left(2 \tilde{\phi}_B-h'\right)+h'' \left(2 h h''-h'^2+4 \tilde{\phi}_B^2\right)\right)}{ h \left(h'-2 \tilde{\phi}_B \right)^2}\label{tmmchi}.
\end{align}
It is easy to see that eq.(\ref{hchi}) now becomes 
a third order non-linear equation for $h$ in the presence of the source $T^m_{--}$  which is difficult to solve. 
To proceed we consider only situations where the infalling matter is varying slowly with $x^-$.  
More precisely, we take the frequency associated with this  variation $\omega$, to satisfy the condition
\be
\label{condfeq}
{\omega \over {\cal J}} \ll \tilde{\phi}_B.
\ee
The reader will note that this is analogous to the condition imposed on the temperature $T$ in the previous section, eq.\eqref{condo}. 
For the kind of situations we consider, in this approximation, only the terms involving the minimum number of derivatives will survive;
in the numerator of eq.(\ref{tmmchi}) the surviving term goes like $4 \tilde{\phi}^2_B h''$, while  in the denominator 
it  goes like $4 \tilde{\phi}_B^2 h$. Retaining these and neglecting the others gives, 
\begin{align}
T_{--}^{\chi}=\frac{N}{12\pi}\frac{h''}{h}\label{tmmchi2},
\end{align}
leading to 
\begin{align}
h'''&=-\zeta \frac{h''}{h}-16\pi G\,T^{m}_{--}\label{finaleom},
\end{align}
where $\zeta$ is defined in eq.\eqref{zeta}.  
This  is a much simpler equation to solve. 

In the  terms involving additional derivatives in eq.(\ref{tmmchi}), each derivative is accompanied by  an  additional factor of $h$; we will self-consistently argue in appendix \ref{latehap} that the condition eq.(\ref{condfeq}) is sufficient to suppress them.


We also note that when eq.(\ref{condfeq}) is met,
\be
\label{valchife}
\chi=\ln z -\ln({h\over 2\tilde{\phi}_B}).
\ee

The formula for the ADM mass, derived   in \ref{amchi}, is given by 
\be
\label{amass}
M={\, \mathcal{J}\over 64\pi G}(h'^2-2 h h'' -2 \zeta h').
\ee
From eq.\eqref{mpropt} one finds that 
\be
\label{dervm}
\partial_{\hat t} M= T^m_{\hat t \hat t}.
\ee
In the absence of infalling matter we see that $M$ is constant. Also for $T^m_{\hat t \hat t}>0$, $M$ increases. 

\subsubsection{Some Additional Comments}
\label{adref}
A few comments are in order here. 

First, the reader might worry that   that the value of $\chi$ we start  with at  $x^-<0$,  before matter begins to fall in,  eq.(\ref{valchia}), is in fact singular at the past Poincar\'e horizon. This follows from noting that 
\be
\label{achi}
\chi=-\omega=\ln z  +\ln{(2\tilde{\phi}_B \mathcal{J})},
\ee
 and $z\rightarrow \infty$ at the horizon. 
It is best to regard this case   as the  limit of the finite temperature situation. For the the eternal black hole, in suitable Poincar\'e coordinates
where  $\phi$ is given by 
\be
\label{valphibha}
\phi={1-\mu x^+x^- \over {\cal{J}} (x^+-x^-)},
\ee
 $\chi$, eq.\eqref{chill}, takes the form
\be
\label{valchiaa}
\chi=\ln(x^+ -x^-\over 2) -\ln(1+\sqrt{\mu}x^+)-\ln(1-\sqrt{\mu} x^-) + \ln(2(\tilde{\phi}_B\mathcal{J}+\sqrt{\mu})).
\ee
This gives a finite value for $\chi$ at the horizon. 
In the limit $\mu\rightarrow 0$ we then get the value for $\chi$ in eq.(\ref{achi}).

Second,  we can, in fact,  start with a black hole of non-zero mass and  redo the analysis of subsection \ref{infallchi} above. 
In this case, for $x^-<0$, $\chi$ and $\phi$ are  given by eq.(\ref{valchiaa}), eq.(\ref{valphibha}) respectively.
In the subsequent evolution 
\be
\label{valchiab}
\chi=\ln{z} -\ln(1+ \sqrt{\mu}x^+) + f_-(x^-),
\ee
and $\phi$ is given by eq.(\ref{gendilq}) since eq.\eqref{finalvaltpp},\eqref{condstress} and eq.\eqref{Tpmanomaly} continue to hold.
It also  follows from the Dirichlet boundary condition $\chi$ satisfies at the boundary that $f(x^-)$ is
\begin{equation}
\label{fmchidi}
f_-(x^-)=-\ln{z_B(x^-)}+\ln(1+\sqrt{\mu}(x^- + 2z_B(x^-))),
\end{equation}
and therefore
\begin{equation}
\chi=\ln{z} -\ln(1+ \sqrt{\mu}x^+)-\ln{z_B(x^-)}+\ln(1+\sqrt{\mu}(x^- + 2z_B(x^-)))\label{chifubh}.
\end{equation}
where $z_B(x^-)$ is given by eq.\eqref{bdytraj}.
Then the `$--$' component of Einstein's equation, which determines $h$ by eq.\eqref{hchi}, using eq.\eqref{fmchidi} and eq.\eqref{tpmexp}, becomes
\begin{align}
h''' &=-\zeta\pqty{\frac{(4\tilde{\phi}_B^2-h'^2+2 h h'')(2\mu h +(1+\sqrt{\mu}x^-)((1+\sqrt{\mu}x^-)h''-2\sqrt{\mu}h'))}{h(2\sqrt{\mu}h+(1+\sqrt{\mu}x^-)(2\tilde{\phi}_B-h'))^2}}\nonumber\\
&-\zeta\pqty{\frac{hh'''(1+\sqrt{\mu}x^-)(2\sqrt{\mu}h+(1+\sqrt{\mu}x^-)({2\tilde{\phi}_B-h'}))}{h(2\sqrt{\mu}h+(1+\sqrt{\mu}x^-)(2\tilde{\phi}_B-h'))^2}}-16\pi G \,T^m_{--}\label{mmhm}.
\end{align}
We can simplify the above equation by using the approximation eq.(\ref{condfeq})
and also taking 
\be
\label{condbs}
{\sqrt{\mu}h\over\phi_B}\ll 1, 
\ee
which follows from eq.(\ref{condo}) if the initial temperature of the black hole we start with is small and $h\le {1\over \cal {J}}$ during the subsequent evolution.  This gives, from eq.(\ref{mmhm}),
\begin{equation}
h'''=-\zeta\pqty{{h''\over h}-{2\sqrt{\mu}\over   1+\sqrt{\mu}x^-}{h'\over h}+{2\mu\over (1+\sqrt{\mu} x^-)^2}}-16\pi G\, T_{--}^m\label{hfbhen}.
\end{equation}
It is straightforward to verify that 
\begin{equation}
h(x^-)= a_1(1+\sqrt{\mu}x^-)+a_2(1+\sqrt{\mu}x^-)^2\label{gnslhb},
\end{equation}
is a solution to eq.\eqref{hfbhen}, for arbitrary constants $a_1$ and $a_2$.
We will show self-consistently  in appendix \ref{latehap},  that  starting with a  black hole which meets the condition, eq.(\ref{condo}) and with slowly infalling matter, which meets eq.\eqref{condfeq}, eq.(\ref{condbs}) is also    valid and  neglecting the additional terms in eq.(\ref{mmhm}) leading to 
eq.(\ref{hfbhen}) can be justified. 
 

Finally, the mass formula, eq.\eqref{amass} obtained above differs from the classical case, eq.\eqref{admclass}. It was mentioned after eq.(\ref{admclass}) that the expression in the classical system can be written in terms of the Schwarzian derivative and therefore preserves $SL(2,R)$ invariance.  It seems somewhat surprising at first therefore that an additional term is present in eq.(\ref{amass}), proportional to $\zeta h'$. In fact this term  is not proportional to the Schwarzian derivative and one might wonder how its presence is consistent with $SL(2,R)$ invariance. 

To understand this better we note that  the starting forms for $\phi$ and $\chi$, eq.(\ref{valphibha}) and eq.(\ref{valchiaa}) are not invariant under a general 
$SL(2,R)$ transformation,
\be
\label{gensl2r}
x^\pm \rightarrow {p x^\pm + q \over r  x^\pm + s}.
\ee
Under such a transformation, while $\phi$ continues to be given by the same form as eq.(\ref{gendilq})  with $h$ transforming as
\be
\label{htransf}
h (x^-) \rightarrow  \frac{ h (x^-)}{( r  x^- +s)^2},
\ee
$\chi$ assumes the form,  
\be
\label{valchiabc}
\chi =\ln  z -\ln(1+ a x^+) + f_-(x^-),
\ee 
with 
\be
\label{vlaaa}
a=   {\sqrt{\mu} s -r\over p -q\sqrt{\mu}}.
\ee
More generally starting from $\chi$ as given in eq.(\ref{valchiabc}), we would get $\chi$ of the same form with
\be
\label{transformationa}
a\rightarrow  {a s -r\over p -q a }.
\ee

 One can then derive an expression for the mass by repeating the analysis above in the more general Poincar\'e coordinates obtained after doing such an $SL(2,R)$ transformation, the details of which are shown in appendix \ref{amchi}, to get
 \be
 \label{bmass}
 M=\frac{\mathcal{J}}{64\pi G}\pqty{h'^2- 2 h h''-2\zeta\pqty{h'-{ 2\,a\,h\over a\,t +1}}}.
 \ee
 We see that there are two additional terms now compared to the classical expression.  Using eq.\eqref{relthatt} we can rewrite this expression in terms of the Poincar\'e time $t$ as, eq.\eqref{minbdt}
 \be
 \label{cmass}
M_\chi=-\frac{1}{8\pi G \mathcal{J}}\,\text{Sch}{(t,\hat{t})}-\frac{N\, \, }{12\pi}\pqty{\frac{t''}{t'}-\frac{2 a\,t'}{a\, t+1}},
 \ee
 where $\hat{t}$ is the FG time coordinate and primes denote derivatives with respect to ${\hat t}$. 
  
Under an $SL(2,R)$ transformation it turns out that the two additional  terms above are together also invariant, as discussed in appendix \ref{amchi}. 
To summarise some of these comments,  an additional parameter, $a$,  enters in determining the mass for the $\chi$ system.
It specifies the    initial conditions for the $\chi$ field and enters in   the second term on the RHS in eq.(\ref{valchiabc}), $\ln(1+a\, x^+)$. 
The presence of this additional parameter allows additional terms to arise in the mass formula, consistent with $SL(2,R)$ invariance. 

We had mentioned above  that the equation of motion in the classical case arises from the Schwarzian action on the boundary coupled to
the  infalling matter stress tensor, eq.(\ref{bait}).  We see from eq.(\ref{hfbhen}) that due to the effects of the $\chi$ field there should be  extra terms in the action; these also cannot 
arise from the Schwarzian derivative term, as in the case of the mass. We have not fully explored this issue but expect that due to the dependence on the extra parameter $a$ the resulting action giving rise to eq.(\ref{hfbhen})  will also  be $SL(2,R)$ invariant. 
\subsection{Detailed Analysis: Quasi Normal Mode}
\label{chidetail}
Let us now return to the case considered in subsection \ref{infallchi} where we start with the Poincar\'e vacuum and  $\chi$ is given in eq.(\ref{valchife}) leading to the equation of motion,   eq.(\ref{finaleom}). While this is  a simpler equation    to solve than the general case, since we have made the approximation, eq.\eqref{condfeq},  it is still quite non-trivial in the presence of  general infalling matter. 

Consider a  situation where matter falls in for some time and then stops. One would expect that eventually the time dependence  dies down and the system  
settles   to a black hole state. This approach to equilibrium is determined by the quasi-normal modes of the system which characterise the final black hole
and is  independent of the details of the initial  infalling matter. The quasi-normal modes describe the ``ring down" of the black hole when subjected to external perturbations.

It turns out that in this system there is only one   quasi-normal mode. 
Let us start with a black hole solution given by eq.(\ref{metic}), eq.(\ref{behdil}), eq.(\ref{chilla}) with mass $M$ given in eq.(\ref{masschi}). 
A coordinate transformation eq.\eqref{coordtr} now brings the metric in  eq.(\ref{metic}) to the Poincar\'e form,
with the dilaton taking the form eq.(\ref{gendilq}) with $h(x^-)$ given by 
\be
\label{vale}
h=h_0={c_1\over \mathcal{J}}(x_0-x^-),
\ee
where 
\be
\label{valc1}
c_1=2{\sqrt{\mu}},
\ee
and
\be
\label{valx0}
x_0={1\over \sqrt{\mu}}.
\ee
This coordinate transformation is being chosen with an eye to the discussion which follows. 
We note that in  these $x^+,x^-$ coordinates the future and past horizons, ${\cal H}^\pm$, lie at $x^+=x_0$ and $x^-=-\infty$ respectively. 



Let us  denote  the dilaton and $\chi$ perturbations in the quasi-normal mode as  $\delta \phi, \delta \chi$ respectively. 
The quasi-normal mode should be regular at the future horizon, $x^+=x_0$. 

In the Poincar\'e coordinates,  
the $\chi$ equation of motion leads to 
\be
\label{constdelchi}
\delta \chi=\delta f_+(x^+)+\delta f_-(x^-).
\ee
Requiring that we start with the  black hole solution and study its ring down leads to the condition that at  ${\cal H}^-$, $ \delta \chi$ vanish, leading to 
$\delta f_+=0$. 

The `++' and `+-'  components of the metric equations then lead to the dilaton being of the form, eq.(\ref{gendilq}), 
with $h$  satisfying the source free equation,($T_{--}^m=0$) eq.\eqref{finaleom},
\be
\label{free}
h'''= -\zeta {h''\over h}.
\ee
 Expanding 
 \be
 \label{expo}
 h=h_0+\delta h,
 \ee
 we then get that the perturbation $\delta h$ satisfies the equation
 \be
\label{perta}
\delta h'''= -\zeta {\delta h''\over h_0}.
\ee
It is easy to see that the solution to this equation  for a black hole of mass $M$ eq.(\ref{masschi}) we started with  is 
\be
\label{apph}
\delta h=-\frac{c_2}{\mathcal{J}(\alpha+2)} (x_0-x^-)^{\alpha + 2},
\ee
where 
\be
\label{valalpha}
\alpha = {\zeta \mathcal{J} \over c_1}.
\ee
and $c_2$ is an arbitrary constant.
From eq.(\ref{gendilq}),{(\ref{valchife})} by considering sufficiently late times i.e., $x^-$ close to $x_0$, and using the transformation to Schwarzschild coordinates, eq.\eqref{coordtr}  it then follows that 
\begin{align}
\delta \phi & \sim{ c_2\over 2\,\mathcal{J}} \,e^{- ({\zeta \mathcal{J}\over 2} + 2\pi T )(t_s-r_*)} \label{pertdila},\\
\delta \chi & \sim  {c_2\over \zeta \mathcal{J}+8\pi T}\, e^{- ({\zeta \mathcal{J}\over 2} +2\pi T )(t_s-r_*)} \label{pertchia}.
\end{align}
The universal exponent $({\zeta \mathcal{J}\over 2} + 2\pi T )$ then characterizes the approach to the black hole solution. 
The quasi-normal mode is given by eq.(\ref{pertdila}) and eq.(\ref{pertchia}).

When $\zeta\rightarrow 0$ it is easy to see from eq.\eqref{apph} that the solution for $\delta h$ corresponds to a black hole solution and therefore no quasi-normal mode exists. The mass corresponding to the perturbation eq.\eqref{apph} can be computed by the linearized form of eq.\eqref{amass} around eq.\eqref{vale} and vanishes as expected.

Let us now consider  a situation where starting with the Poincar\'e vacuum matter begins to  fall in at $x^-=0$ and stops after some time.
We take $T^m_{--}>0$ during the infall. At $x^-=0$, $h'=h''=0$, since for $x^-<0$, $h$ is given by eq.\eqref{meh} 
It follows from eq.\eqref{finaleom} then that $h''$ must be negative for $x^->0$,  even after the matter stops falling in, and thus $h'$ must also be negative for $x^->0$ as discussed  in appendix \ref{latehap}.  As a result $h$ will monotonically decrease and eventually vanish, say  at $x^-=x_0$ . Let us consider a case where $h>0$ when the matter stops falling in. In the subsequent evolution the mass is conserved and $h$ satisfies the source free equation, eq.\eqref{finaleom}, with $T^m_{--}=0$. Thus  from eq.(\ref{vale}) and (\ref{apph}) we know that in the vicinity of $x_0$ 
\be
\label{valah}
h={c_1\over \mathcal{J}}(x_0-x^-)  -\frac{c_2}{\mathcal{J}(\alpha+2)} (x_0-x^-)^{\alpha +2}.
\ee
 and the system will ``ring down" to the   black hole corresponding to the  final mass which determines $c_1$ in terms of mass M by eq.(\ref{valc1}), eq.\eqref{masschi}.
 
 More details illustrating this behaviour are given in appendix \ref{latehap}. 
 
 We also note, as was discussed before in section \ref{infall}, that where $h\rightarrow 0$ in a generic way with a first order zero, boundary time ${\hat t}\rightarrow \infty$. Situations where this happens while matter is falling in, are less universal and need to be analysed on a case-to-case basis. 
 
 Let us end this subsection with some final   comments. In the analysis above we took $\chi$ to be of the form, eq.\eqref{chisol2} which is  appropriate if we are starting with the Poincar\'e vacuum.  Instead, if we started with a black hole of non-zero mass, the form $\chi$ takes is eq.(\ref{valchiab}). 
 One can then repeat the quasi-normal mode analysis.  In this case $h$ satisfies  eq.\eqref{hfbhen}, with the  solution,
 \begin{equation}
 h={1\over \mathcal{J}}\pqty{1-\mu (x^-)^2}.\label{hbhsolu}
 \end{equation}
 which is obtained for the choice $a_1=2$ and $a_2=-1$ in eq.\eqref{gnslhb}. Denoting this solution as $h_0$ and considering a small fluctuation about it $\delta h$, eq.\eqref{expo}, we get that $\delta h$ satisfies the equation 
 \be
 \label{perch}
 \delta h'''=-\zeta\pqty{\frac{2\mu}{(1+\sqrt{\mu}x^-)^2}\frac{\delta h}{h_0} - \frac{2\sqrt{\mu}}{1+\sqrt{\mu}x^-}\frac{\delta h'}{h_0} + {\delta h''\over h_0}}.
 \ee
General solution of the above equation for $\delta h$ can be written as a sum of three independent solutions
  as: 
 \be
 \label{gens-perch}
 \delta h = k_1 \delta h_1  + k_2 \delta h_2+ k_3 \delta h_3,
 \ee
 where 
 \begin{align}
 \delta h_1 & =  1+\sqrt{\mu} x^- \label{phi11a}, \\
 \delta h_2 & = (1+\sqrt{\mu} x^-)^2 \label{phi11b}, \\
 \delta h_3 & = \left(1-\sqrt{\mu } x^-\right)^{\frac{  \zeta \mathcal{J} }{2 \sqrt{\mu }}+2} \left(1+\sqrt{\mu } x^-\right)^{-\frac{\zeta \mathcal{J} }{2 \sqrt{\mu }}} \label{phi11c},
 \end{align}
 and $k_1, k_2,k_3$ are arbitrary constants. 
 Of these the first term can be shown to change the black hole mass, see eq.\eqref{gnslhb} 
  and therefore  cannot not arise at late times. The second term, although does not alter the mass, can be set to zero by an $SL(2,R)$ transformation. This leaves only the third solution eq.(\ref{phi11c}) which at late times, $x^- \rightarrow \frac{1}{\sqrt{\mu}}$, becomes
  \begin{equation}
  \delta h_3\sim (1-\sqrt{\mu}x^-)^{2+{\zeta \mathcal{J} \over 2\sqrt{\mu}}}\label{dhlte}.
  \end{equation}
This is the quasi-normal mode   with   an exponent which agrees with what was obtained earlier, eq.\eqref{apph},\eqref{valalpha}.
 In fact it is easy to see that for this solution, at late time, only the last term on the RHS of eq.(\ref{perch}) contributes so that the perturbation satisfies eq.(\ref{perta}) obtained earlier.

 This shows that the exponent $({\zeta \mathcal{J} \over 2}+ 2 \pi T)$ which characterises the quasi-normal mode appears quite universally, regardless of the
 initial conditions for $\chi$, as would be expected on physical grounds.  
 
 \subsection{Second Law}
 \label{chisecondlaw}
 In this subsection we show that the entropy $S^{\chi}_{gen}$ monotonically increases along the event horizon. The entropy is defined at the event horizon in  eq.\eqref{entropchi}, and includes a contribution due to the 
 $\chi$ field. We saw in section \ref{chitherm} that with this definition the First law is satisfied. 
 Here we consider non-equilibrium situations  with   infalling matter of the kind that was considered in section \ref{infall}. We will find that the second law is true as long as the energy
 condition, eq.\eqref{energycond} is satisfied. 
 
 We will restrict ourselves to situations where the matter stops falling in after some time. We saw that the solution is determined by the function $h$
 which at late times $x^-\rightarrow x_0$,  takes the form, eq.(\ref{valah}).  The event horizon, where $(\nabla \phi)^2=0$, at late times, is given by 
 \be
 \label{evenh}
 x^+=x_0.
 \ee

 From  eq.\eqref{gendilq} and using the form for $\chi$ eq.\eqref{chisol2} we see that the entropy is, 
 \begin{align}
S^{\chi}_{gen}&=\pqty{\frac{\phi}{4G}-\frac{N\chi}{6}}\Big\vert_{eh}\nonumber\\
&=\frac{1}{4G}\pqty{{h\over x_0 - x^-}+{h'\over 2}+\frac{\zeta}{4}-{\zeta \over 2}\pqty{\ln{z}-\ln{\pqty{\frac{h}{2\tilde{\phi}_B-h'}}}}}\Big\vert_{eh}\label{entchi}.
\end{align}
Note we have not assumed that the infalling matter is slowly varying with eq.(\ref{condfeq}) being met and our analysis below will apply to the general case.

 To examine the second law, we proceed similar to the classical case discussed in section \ref{secondclass}. 
 The affine parameter along the event horizon is given by a relation analogous to eq.(\ref{affine}) given by 
 \begin{equation}
 {\text{d}x^-\over \text{d}\lambda}=(x_0-x^-)^2= 4\, e^{-2\omega}\Big\vert_{eh}\label{affchi}.
 \end{equation}
  From eq.\eqref{entchi}, we get 
\begin{align}
\frac{\text{d}S^{\chi}_{gen}}{\text{d}\lambda}&= (x_0-x^-)^2\nabla_{-}S^{\chi}_{gen}\nonumber\\
&=\frac{(x_0-x^-)^2}{4G}\pqty{\frac{h}{(x_{0}-x^{-})^{2}}+\frac{h'}{x_{0}-x^{-}}+\frac{h''}{2}+\frac{\zeta}{2}\pqty{\frac{1}{x_{0}-x^{-}}+\frac{h'}{h}+\frac{h''}{2\tilde{\phi}_B-h'}}}\label{dels}.
\end{align}
At late time,  $x^-\rightarrow x_0$, $h$ is given by eq.\eqref{vale} and we see that 
\begin{align}
\frac{\text{d}S^{\chi}_{gen}}{\text{d}\lambda}&\rightarrow 0 \label{delsbeh}.
\end{align}
Let us pause to make one comment here. 
We took $\chi$ as given in eq.(\ref{chisol2}) in obtaining eq.(\ref{entchi}). 
More generally, starting with a black hole, instead of the Poincar\'e vacuum, $\chi$ is given by eq.(\ref{valchiab}), with $f_-(x^-)$ in turn being  expressed in terms of $h$ by eq.\eqref{fmchidi}. 
In this case one can show that at late time $x^-\rightarrow x_0$, $h$ is given by a  quadratic function, 
\be
\label{genquad}
h= \tilde{k}_1 (1+ \sqrt{\mu} x^-) (x_0-x^-),
\ee
for some constant $\tilde{k}_1$. Repeating the analysis above then shows that eq.(\ref{delsbeh}) is valid in this more general situation as well. 

Next we  compute the second derivative of the entropy, 
\begin{align}
\frac{\text{d}^{2}S^{\chi}_{gen}}{\text{d}\lambda^{2}}=16\,e^{-2\omega}\del_-(e^{-2\omega}\del_-S_{gen}^{\chi})\bigg\vert_{eh}\label{2dsl}.
\end{align}

From the `$--$' component of the Einstein equation, eq.(\ref{geneqs}) and noting that when the $\chi$ field is present, the stress tensor  component $T_{--}$, including $T_{--}^{\chi}$ eq.(\ref{Tppchi}), is given by,

\begin{equation}
T_{--}=\frac{N}{12\pi}\left(-\partial^{2}_{\pm}\chi+\partial_{\pm}\chi\partial_{\pm}\chi+2\del_{\pm}\chi\del_{\pm}\omega\right)+ T^m_{--}\label{totalstress},
\end{equation}

we get
\begin{equation}
e^{2\omega}\del_{-}\pqty{e^{-2\omega}\del_{-}(\phi-{\zeta\over 2}\chi)}=-\frac{\zeta}{2}(\partial_{-}\chi)^{2}-8\pi G \,T^m_{--}. \label{mme}
\end{equation}
Using eq.(\ref{mme}) in eq.\eqref{2dsl} we see that 
\be
\label{final}
\frac{\text{d}^{2}S^{\chi}_{gen}}{\text{d}\lambda^{2}}=\frac{4\,e^{-4\omega}}{G}\pqty{-\frac{\zeta}{2}(\partial_{-}\chi)^{2}-8\pi G \,T^m_{--}}\bigg\vert_{eh}<0,
\ee
where we have used the energy condition eq.\eqref{energycond}.
Since eq.(\ref{delsbeh}) is met as $x^-\rightarrow x_0$, we conclude that 
\be
\label{slaw}
{ \text{d}S^{\chi}_{gen} \over \text{d}\lambda} >0,
\ee
showing that the generalised entropy monotonically  increases along the future event horizon. 

Let us  note here that $x^-\rightarrow x_0$ corresponds to the ``far future" since boundary proper time ${\hat t}\rightarrow \infty$ in this limit, eq.\eqref{bdytime}. 

Before concluding this section let us note that unlike the classical case considered in section \ref{clasappar}, in general the analogous statement for the apparent horizon, defined by eq.(\ref{condapph}), is not valid here, i.e. $S^{\chi}_{gen}$  now  evaluated at the apparent horizon need not increase. We show this by considering an explicit example in appendix \ref{cldlt}. In contrast, when we consider the future Q-screen, which we will discuss in section \ref{qscreens}, the generalised entropy does increase.

\section{Dynamical system with $\psi$ fields}
\label{psi}
Having analysed the $\chi$ system in the previous section we now go back to the starting Lagrangian, eq.\eqref{acts}, and consider the system of $N$ scalars directly 
in the semi-classical limit, $N\rightarrow \infty$, $G\rightarrow 0$, keeping $G N$ fixed. 
We take the $\psi_i$ fields to satisfy Dirichtlet boundary condition
\be
\label{db}
\psi_i |_B=0,
\ee
 at the boundary, eq.\eqref{bdrydil}, as was mentioned above.  
 Let us consider a boundary described in conformal coordinates by the trajectory,
 \be
 \label{traj}
 x_B^+=x^+(x^-).
 \ee
 And consider the $x^+$ modes to be in the vacuum with respect to a coordinate $x_{v}^+$.
 Then expanding the $\psi_i$ fields as 
 \be
 \label{exppsi}
 \psi=\int_{0}^\infty d\omega \Big[{a_+(\omega) \over \sqrt{2 \omega}} e^{-i\omega x_{v}^+} + f(\omega, x^-) + c.c.\Big],
 \ee
 and imposing boundary conditions, eq.(\ref{db}) gives 
 \be
 \label{exppsi2}
 f(\omega, x^-)=-{a_+(\omega)\over \sqrt{2 \omega}} e^{-i\omega x_v^+(x_B^+)}.
 \ee
 This shows that the $x^-$ modes of $\psi_i$ fields will be in the vacuum with respect to a coordinate
 \be
 \label{coordxm}
 x_v^-=x_v^+(x_B^+).
 \ee
 
We also note that the conformal anomaly is given by 
\begin{equation}
\label{confanom}
    T_{+-}=-\frac{N}{12\pi}\del_{+}\del_{-}\omega.
\end{equation}
Substituting in the `$+-$' component of the metric equation of motion, eq.\eqref{pm} we get that 
\be
\label{eqphtilde}
2\partial_+\partial_-{\tilde \phi}+ e^{2 \omega} {\tilde \phi}=0,
\ee
where 
\be
\label{deftildephi}
{\tilde \phi}= \phi -{GN \over 3}.
\ee

\subsection{Entanglement Entropy}
\label{entent}
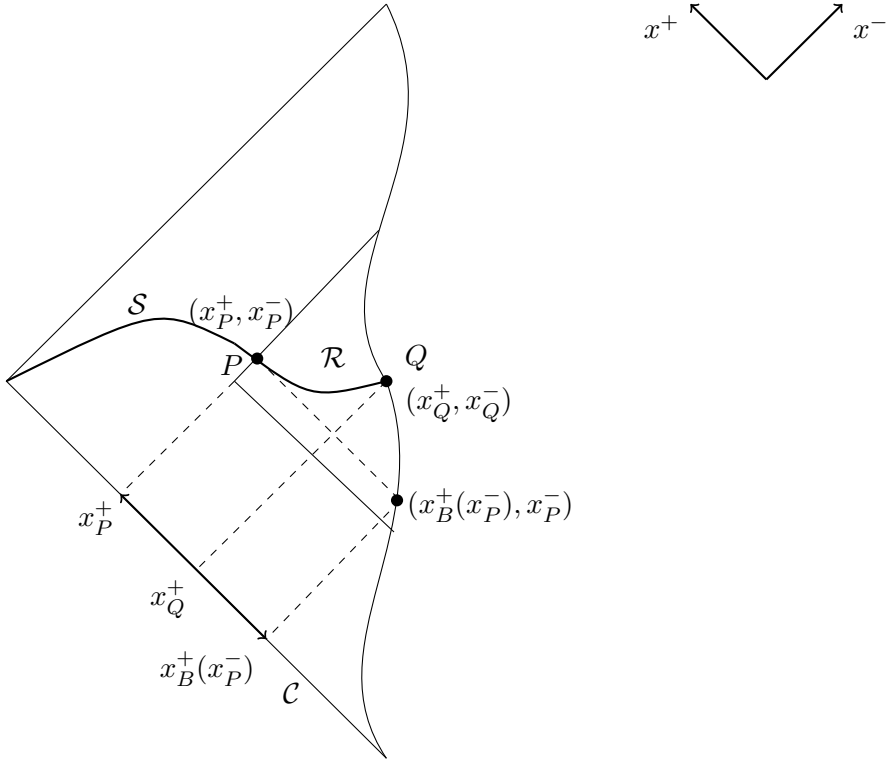
\begin{figure}
	\centering
	\begin{tikzpicture}
	\draw (5,0)  .. controls (4,1.5) and (5.7,3) .. (5,5);
	\draw (5,5)  .. controls (4,6.5) and (6,8) .. (5,10);
	\draw (5,10)-- (0,5) -- (5,0) node[ anchor=south
	east] at (4,0.6) {$\mathcal{C}$};
	\draw (5.1,3) -- (3,5);
	\draw (3,5) -- (4.9,7);
	\draw [thick] (0,5) .. controls (2,6) .. (3,5.5) node[ anchor=south
	east] at (2,5.75) {$\mathcal{S}$};
	\draw [thick] (3,5.5) .. controls (4,4.75) .. (5,5) node[ anchor=south
	west] at (5.1,5) {$Q$};
	\draw	node[ anchor=north 	west] at (5.11,5.1) {$(x_Q^+,x_Q^-)$};
	\draw [dashed] (5,5) -- (2.5,2.5) node[ anchor=north east
	] at (2.5,2.5) {$x^+_Q$};
	\draw node[ anchor=south west] at	(4.0,5.05) {$\mathcal{R}$};
	\fill[black] (5,5) circle (0.08cm);
	\draw [dashed] (3,5) -- (1.5,3.5) node[ anchor=south east] at
	(3.25,4.95) {$P$};
	\fill[black] (3.3,5.3) circle (0.08cm);
	
	\draw  node[ anchor=south west] at	(2.25,5.55) {$(x^+_P,x^-_P)$};
	\draw  node[ anchor=north east] at	(1.55,3.55) {$x^+_P$};
	\draw  node[ anchor=north east] at	(3.4,1.55) {$x^+_B(x^-_P)$};
	\draw [dashed] (3.3,5.3) -- (5.18,3.42);
	\fill[black] (5.14,3.42) circle (0.08cm);
	\draw [dashed] (5.18,3.42) -- (3.42,1.58) ;
	\draw  node[ anchor=north west] at (5.14,3.72) {$(x^+_{B}(x^-_P),x^-_P)$};
	\draw [thick,<->] (1.5,3.5) -- (3.42,1.58) ;
	
	\draw [thick,->](10,9) -- (9,10) node[ anchor=north east] at	(9,10) {$x^+$};
	\draw [thick,->](10,9) -- (11,10) node[ anchor=north west] at	(11,10) {$x^-$};
	\end{tikzpicture}
	\caption{Entanglement of matter fields across the horizon}
	\label{eefig}
\end{figure}

The entanglement entropy due to the scalar fields $\psi_i$ can be obtained  using the results in \cite{fiola94,Holzhey}.
Consider a single minimally coupled scalar $\psi$ and a space-like slice denoted as ${\cal S}$ in Fig \ref{eefig} which intersects the boundary at the point $Q$.
We are interested in the entanglement entropy of the region extending from a point $P$ to $Q$ along  ${\cal S}$. We will denote this region as ${\cal R}$ below. 
The entanglement entropy  depends on the state of $\psi_i$. In the presence of the boundary the state of the left-moving modes 
($x^-$ dependent) is determined by the state of the right-moving modes ($x^+$ dependent) and the location of the boundary, see eq.(\ref{coordxm}). Therefore it is enough to specify the state of the right-moving modes. We take this state to be in the vacuum with respect to the $x_v^+$ coordinate.  
We will denote the coordinate of $P,Q$ as $x^\pm_P$, $x^\pm_Q$ below. Now consider an initial  null slice ${\cal C}$, the state of the left-movers can be specified on it. The right-moving modes present in region ${\cal R}$ of ${\cal S}$ correspond to right-movers on ${\cal C}$ that lie in the interval 
 $[x^+_P, x_Q^+]$ . The left-movers in ${\cal R}$ correspond to right-movers that lie in the interval $[x_Q^+,x_B^+(x_P^-)]$ which after reflecting off the boundary have turned into left-movers. 
 The entropy in ${\cal R}$ is therefore the entanglement in the right-moving modes lying in the region $[x^+_P, x_Q^+] \cup [x_Q^+,x_B^+(x_P^-)]$ 
 on ${\cal C}$. 

It follows then from   \cite{fiola94} that this  entanglement entropy is given by 
 
 \be
 \label{Senta}
  S_{EE}={1 \over 12} \left[\ln{(\Delta x_v^+)^2 \over \delta^2}+ 2 \rho_P \right].
  \ee
 Let us explain the different terms on the RHS above. 
 $\Delta x_v^+$ is the length of the interval $[x^+_P, x_Q^+] \cup [x_Q^+,x_B^+(x_P^-)]$  on ${\cal C}$ measured in the vacuum coordinate $x_v^+$. 
 That is, 
 \be
 \label{valhelxp}
 \Delta x_v^+=x_v^+(x_P^+)-x_v^+(x_B^+(x_P^-)).
 \ee
 $\delta ^2$ in eq.(\ref{Senta}))  is an invariant cut-off which needs to be introduced to obtain a finite result. Finally,  the metric in the coordinates $x_v^\pm$ is given by 
 \be
 \label{defrho}
 \text{d}s^2=-e^{2\rho} \text{d}x_v^+ \text{d}x_v^-,
 \ee
 and $\rho_P$ is the conformal factor at the point $P$ in the $x_v^\pm$ coordinates
 
 The result eq.(\ref{Senta}) does not depend on the choice of the initial surface ${\cal C}$. Also, the vacuum state of the system is only specified up to an $SL(2,R)$ transformation, since the notion of positive frequency modes does not change under such a transformation. 
 It is easy to check that under such a transformation
 \be
 \label{sl2rp}
 x_v^+\rightarrow {a x_v^++ b\over cx^+_v +d},
 \ee
 with $ad-bc=1$,
 $x_v^-$ given by eq.\eqref{coordxm} also transforms under the same $SL(2,R)$ transformation
 \be
 \label{sl2rm}
  x^-_v\rightarrow {a x^-_v+ b \over c x^-_v+d},
  \ee
  and the entanglement entropy eq.(\ref{Senta}) is invariant. 
 
 We will use eq.(\ref{Senta}) in the discussion below often with the point $P$ lying on the horizon of black hole. Also, we will take all $N$ scalar fields to be in the same state. The full entanglement entropy will then be multiplied by a factor of $N$ and is given by 
 \be
 \label{SentaN}
  S_{EE}={N \over 12} \Big[\ln{(\Delta x_v^+)^2 \over \delta^2}+ 2 \rho_h\Big],
  \ee 
 where $\rho_h$ denotes the conformal factor at the horizon.

\subsection{Thermodynamics}
\label{thermpsi}
We are now ready  to  study black holes in this system, our analysis parallels that of section \ref{chitherm} in the $\chi$ case. 
The black hole  solution is given by the metric   in eq.(\ref{metic}) and the dilaton given in eq.(\ref{behdil}). We will show that this solution continues to solve the equations of motion in the $\psi$ system as well, for a suitable choice of vacuum for the matter fields. 
We will use  both  the conformal coordinates, $x_s^\pm$ given in eq.(\ref{defxpmther}) 
and Kruskal coordinates, $X_K^\pm$ given in eq.\eqref{poinkrus} in the discussion below. 
We will also sometimes refer to the  $x^+, x^-$ directions as ``right-moving and left-moving " coordinates respectively

We take the $\psi_i$ fields to be in the vacuum with respect to $X_K^\pm$ coordinates. 
It then follows that the `$++$' and `$--$' components of the stress tensor  in the $x_s^\pm$ coordinates are given by (see appendix \ref{appcoord}),
\begin{align}
 T^{\psi}_{++}&= -\frac{N}{24\pi}\text{Sch}(X_K^+,x_s^+), \label{ppm} \\
 T^{\psi}_{--}&= -\frac{N}{24\pi}\text{Sch}(X_K^-,x_s^-), \label{mmm}
 \end{align}
and agree with the values they take in the $\chi$ system, eq.(\ref{Tpp}). 
This shows that the solution eq.(\ref{metic}) and eq.(\ref{behdil}) satisfies the equations of motion. 

The temperature for the solution eq.(\ref{metic}) is given by eq.(\ref{valT}).  The mass for the black hole in the $\psi$ system can be calculated as discussed in appendix \ref{mpsi}. 
In the limit eq.\eqref{condo}
the mass is given by 
\begin{equation}
 M=\frac{1}{16\pi G\,\mathcal{J}}\pqty{1-\frac{\zeta}{2\tilde{\phi}_B}}\mu\label{metern}.
 \end{equation}
 Henceforth, we will work in the region of parameter space where 
 \be
 \label{parasp}
 1-{\zeta \over 2 \tilde{\phi}_B} >0.
 \ee
 
We see from eq.\eqref{valmass}, eq. \eqref{masssch1} and \eqref{metern} that the effect of the scalar fields can be incorporated by changing the coefficient of the Schwarzian action. 
One way to do this is to take 
\be
\label{shiftG}
G\rightarrow  {G \over {1-{\zeta \over 2 \tilde{\phi}_B } }  }   ,
\ee
in eq.\ref{valmass}. 
In contrast in the $\chi$ system, the quantum effects result in a contribution to $M$  which goes like $\sqrt{\mu}$, eq.\eqref{masschi}, and thus here the effects cannot be absorbed 
into a renormalisation of the coefficient of the Schwarzian action. 

Let us now calculate the generalised entropy at the horizon. 
This is given by 
\be
\label{genet}
S^{\psi}_{gen}=S_{BH}  + S_{EE},
\ee
where $S_{BH}$,  defined in eq.\eqref{defentbh}, is the classical contribution and $S_{EE}$ is the entanglement in the matter fields  outside the horizon.

From eq.\eqref{behdil} and noting that the location of the horizon is given by eq.\eqref{horizon}, we see that 
\be
\label{dilip}
\phi_h = {\sqrt{\mu}\over \mathcal{J}}+\frac{\zeta}{4},
\ee
so that 
\be
\label{vallabh}
S_{BH}=\frac{1}{4G}\pqty{\frac{\sqrt{\mu}}{\mathcal{J}}+\frac{\zeta}{4}}.
\ee

In evaluating the entanglement entropy we need to take into account  a  subtlety. We took both the left-moving and right-moving modes to be in the Kruskal vacuum above. Let us, to begin, take the right-moving modes to be in the Kruskal  vacuum, then 
since  dilaton eq.\eqref{behdil} results in the boundary $\phi=\phi_B$ satisfying the equation,
\be
\label{bounded}
X_k^+X_k^-=-D,
\ee
where 
\be
\label{defD}
D=\pqty{\tilde{\phi}_B\mathcal{J}-\sqrt{\mu}\over\tilde{\phi}_B\, \mathcal{J}+\sqrt{\mu}},
\ee
it follows from eq.\eqref{coordxm} and eq.\eqref{bounded} that the left-moving modes  would be in the vacuum with respect to the coordinate
\be
\label{vac-}
X_{v}^-=-{D\over X_k^-}.
\ee
This is of course related to $X_K^-$ by an $SL(2,R)$ transformation and therefore $T^{\psi}_{--}$  will agree with 
eq.(\ref{mmm}) above. However in computing the entanglement entropy one  needs the value of the conformal factor $\rho_h$, eq.(\ref{SentaN}) in the coordinate system in which the matter fields are in the vacuum . For the choice we are making here $\rho_h$ must therefore be computed in the 
$(X_K^+, X_v^-)$ coordinate system and not in the $(X_K^+,X_K^-)$ coordinate system which would give a wrong result.

Once we keep this subtlety in mind the rest of the calculation is straightforward. 
It is convenient to calculate the entropy at the future (or past) horizon rather than at the bifurcate horizon. 
At the future horizon $X_K^+ = 0$, (see eq.\eqref{poinkrus}),  one gets that $\rho_h$, eq.\eqref{defrho}, using eq.\eqref{poinmet2} and noting eq.\eqref{vac-}, is given by 
 \be
 \label{valrhovac}
 \rho_h =\pqty{-\ln{(1+ X^+_KX^-_K)\over 2}-\frac{1}{2}\ln{\frac{D}{(X_K^-)^2}}}\bigg\vert_{h} = -{1\over 2} \ln {D\over  4(X_K^-)^2}.
 \ee 
 From Fig \ref{eefig} we see that the interval $\Delta x_v^+$, eq.(\ref{valhelxp}), is given by 
 \be
 \label{valexpa}
 \Delta x_v^+=(X_K^+ - X_v^-)\vert_{h}= \frac{D}{X_K^-}.
 \ee
 Thus the entanglement entropy for the $N$ fields becomes
 \be
 \label{entera}
 S_{EE}= \frac{N}{12}\ln  \pqty{{\frac{4(\tilde{\phi}_B \,\mathcal{J}-\sqrt{\mu})}{\tilde{\phi}_B\, \mathcal{J}+\sqrt{\mu}}}}.
 \ee
 In the approximation, eq.\eqref{condo}, the temperature dependent part of this becomes,
 \be
 \label{enter}
 S_{EE}\approx -\frac{N}{6}\frac{\sqrt{\mu}}{\tilde{\phi}_B\, \mathcal{J}}.
 \ee
 As a result, from \eqref{genet},eq.\eqref{vallabh} and eq.(\ref{enter}) the generalised entropy is given by 
 \be
 \label{genenta}
 S_{gen}^\psi=\frac{1}{4G\,\mathcal{J}}\pqty{1-{\zeta\over 2\tilde{\phi}_B}}\sqrt{\mu}.
 \ee
 
 It is easy to see from eq.\eqref{genenta},  \eqref{metern}  and eq.\eqref{valT} that the first law of thermodynamics is met by the generalised entropy at the event horizon, i.e.,
 \be
 \label{genfl}
 T\text{d}S^{\psi}_{gen}=\text{d}M.
 \ee
 
 Let us end this discussion of the thermodynamics with some comments.
 First we see that  in the $\psi$ system  $S^{\psi}_{gen}$ is always positive,  
 \be
 \label{sph}
 S^{\psi}_{gen}  >0.
 \ee
 In contrast for the $\chi$ system it follows from eq.\eqref{entbchi} and eq.\eqref{valT} that the temperature dependent terms in the entropy are
 \be
 \label{sienna}
 S^{\chi}_{gen}=\frac{\pi \, T}{2G\,\mathcal{J}}  + \frac{N}{6} \ln({T\over \mathcal{J} \tilde{\phi}_B}),
 \ee
 and therefore at a sufficiently low $T$, $S^{\chi}_{gen}$ becomes negative probably indicating that the system no longer behaves in a sensible manner.  
 
 Second, in the discussion above in the $\psi$ system we started with the right-movers being in the Kruskal vacuum; instead if we had 
 taken them to be in an $SL(2,R)$ transformed vacuum and changed the left-movers  ground state also consistently under the same $SL(2,R)$ transformation from the vacuum state we started with  above, eq.(\ref{vac-}), then the entanglement and hence generalised entropy would  have  remained unchanged from our result above, eq.(\ref{enter}).  
 This follows from the  comments towards the end of section \ref{entent}. 
 Finally, one can show that the metric eq.(\ref{metic}) and dilaton eq.(\ref{behdil}) satisfy the equations of motion even when the approximation eq.\eqref{condfeq} is not true, as long as the left and right-movers are in the Kruskal vacuum (or vacua related to them by an $SL(2,R)$ transformation)\footnote{This follows from the fact that eq.(\ref{behdil}) can be written in the form eq.(\ref{gendilq}) in Poincar\'e coordinates, with $h=\frac{1}{\mathcal{J}}(1-\mu (x^-)^2)$, so that $h'''=0$; it then follows from eq.\ref{fulltpsi} that in Poincar\'e coordinates $T_{--}^\psi=0 $. }.
 This means that in general $S^{\psi}_{gen}$ is given by eq.(\ref{entera}).
 If we now assume the first law, we can calculate the mass of the system in this more general situation  to be 
 \be
 \label{apparmass}
 M=\frac{\mu}{16\pi G\,\mathcal{J}}+{N\tilde{\phi}_B \,\mathcal{J}\over 24\pi^2}\ln(\tilde{\phi}_B^2\,\mathcal{J}^2-\mu). 
 \ee

 \subsection{Infalling matter}
 \label{psilaws}
 We now couple the system to an additional classical massless field and consider a situation where this  massless field is purely infalling (left-moving) so that its  stress tensor, which we denote as  $T^m_{\mu\nu}$, has only one non-zero component, $T^m_{--}$. Instead of adding this extra field we could have considered a situation where 
 there was an infalling  coherent state made up of  the $\psi_i$ matter fields themselves. This situation can also be analysed along the lines below, but has some additional complications which we would rather avoid here.  
 
  We  choose the initial state for the $\psi_i$  fields to be   the vacuum in the Poincar\'e $x^+$ coordinate for the right-moving  modes.   The state for the left-moving modes is then given by eq.(\ref{coordxm}) to be the vacuum with respect to the coordinate 
 \begin{equation}
x_{v}^{-}=x^{-}+\frac{2h}{{2\tilde{\phi}_{B}}-h'}  \label{vac}.
\end{equation}
Noting that in the Poincar\'e vacuum $\langle x|T_{--}|x\rangle$, which is the first term on the RHS of eq.(\ref{strtran}),
 vanishes,  we see that the stress tensor component $T^\psi_{--}$  in the vacuum state eq.\eqref{vac} is  given by 
\begin{align}
T_{--}^{\psi}&=-\frac{N}{24\pi}\text{Sch}(x_{v}^{-},x^{-}) \label{Tpsi}.
\end{align}
 Using eq.\eqref{vac} we get
\begin{equation}
T_{--}^{\psi}=-\frac{N}{24\pi}\pqty{\frac{-6 h^2 h'''^2+2 h h'''' \left(2 h h''-h'^2+4 \tilde{\phi}_B^2\right)+4 h''' \left(h'+\tilde{\phi}_B\right) \left(2 h h''-h'^2+4 \tilde{\phi}_B^2\right)}{\left(-2 h h''+h'^2-4 \tilde{\phi}_B^2\right)^2}}\label{fulltpsi}.
\end{equation}
Again, we consider the slowly varying limit eq.\eqref{condfeq}. It this limit  we get
\begin{equation}
T_{--}^{\psi}=-\frac{N}{24\pi}\frac{h'''}{\tilde{\phi}_B}\label{ctpsi}.	
\end{equation}
It then follows that the equation for $h$  which is 
\be
\label{her}
h'''=-16\pi G\,T_{--}^{\psi}-16\pi G\,T^{m}_{--},
\ee
becomes 
\be
\left(1-\frac{\zeta}{2\tilde{\phi}_B}\right) h'''=-16\pi G T_{--}^m\label{heqwllj},
\ee
where $\zeta$ is given by eq.\eqref{zeta}.

We see in comparison with the classical case, eq.\eqref{mmcim}, that  as in the discussion of the mass in eq.\eqref{metern} above,  the quantum effects of matter can be incorporated by 
changing the coefficient of the Schwarzian action, or equivalently by rescaling $G$ as given in eq.\eqref{shiftG}. 

  It then follows that $h$ is given in terms of $T_{--}^m$ by eq.(\ref{gensolaa}) and the mass by eq.(\ref{admclass}) after rescaling $G$ in subsection \ref{infall}
 by eq.\eqref{shiftG}. Starting with the Poincar\'e vacuum, consider the situation where matter starts falling in at $x^-=0$ and stops at $x^-=x_f$. Denoting 
 \begin{equation}
 \mu=\frac{8\pi G\,\mathcal{J}}{\pqty{1-\frac{\zeta}{2\tilde{\phi}_B}}}\int_{0}^{x_f} T^m_{--}(x^-) \,\text{d}x^-,\label{mudefinf}
 \end{equation}
we find that $h$, for $x^->x_f$, is given by
\begin{equation}
h(x^-)=\frac{1}{\mathcal{J}}-\frac{2}{\mathcal{J}}\int_{0}^{x^-}\pqty{\int_{0}^{x^-}\mu\,\text{d}x^-}\text{d}x^- =\frac{1}{\mathcal{J}}\pqty{1-\mu (x^-)^2}\label{hinchi}
\end{equation}
The expression for the dilaton with this $h(x^-)$  is given by eq.\eqref{dil1}.
The mass can then be calculated using eq.\eqref{psiinf} and is given by eq.\eqref{mrespsi}.

Using the coordinate transformations mentioned in appendix \ref{appcoord}, we find that the metric, eq.\eqref{poin} and the dilaton, eq.\eqref{dil1},  transform to the form given by eq.\eqref{bhmet3} and eq.\eqref{dilr} respectively. It then follows that the temperature of the black hole corresponding to eq.\eqref{hinchi} is given by the expression eq.\eqref{valT}, where $\mu$ is given by eq.\eqref{mudefinf}. Also, the entropy will be given by the eq.\eqref{genenta}. It then follows immediately that the system obeys the first law of thermodynamics, $T\text{d}S_{gen}^{\psi}=\text{d}M$, which shows that the system equilibriates instantly after the matter ceases to fall in. It is easy to see from eq.\eqref{hinchi}, that $h$ vanishes at late times, $x^-\rightarrow\frac{1}{\sqrt{\mu}}$, as a linear zero, i.e.,
\begin{equation}
h(x^-)\sim \frac{2}{\mathcal{J}}({1}-\sqrt{\mu}x^-).\label{hformp}
\end{equation}

%
%
%
 Let us end this subsection with some comments. In the $\chi$ system we found a quasi-normal mode by studying small fluctuations around the black hole background. Here
 interestingly, it is easy to see  from eq.\eqref{heqwllj}  that once   the matter stops falling in, the system instantaneously equilibriates to a black hole  of the final mass. Thus, there is no non-trivial ringing down and therefore no quasi-normal mode is present. 
  
 If we had started with a black hole one would expect the $x^+$ modes to be in the Kruskal vacuum
instead  of the Poincar\'e vacuum, which we considered as the initial state above. However, it is a curious fact that 
the Kruskal and Poincar\'e  coordinates  are related by an  $SL(2,R)$ transformation in this system, and therefore the two vacua are the same.

Finally, we had mentioned in the introduction that the differences in the  $\chi$ and $\psi$ systems arise due to different boundary conditions. 
The $T_{+-}$ component of the stress tensor agree in the two cases by construction. We  also see from eq.(\ref{Tppmm}), and eq.(\ref{valchiabc})  that  for the $\chi$ system, $T_{++}$ vanishes, which is also true for the $\psi$ system if we start with the Poincar\'e or Kruskal vacua. Thus the differences in the two cases arise due to differences in $T_{--}$ which in turn  differs due to the different boundary conditions.

   \subsection{Second law}
   \label{psisecondlaw}
In this subsection we examine the second law for this system. 
 To examine the second law, we proceed in a manner similar to section \ref{chisecondlaw} and consider the first and second derivatives of the generalised entropy, $S^{\psi}_{gen}$, with respect to the affine parameter $\lambda$, eq.\eqref{affine}, along the event horizon. We again take the initial state for the $\psi_i$  fields to be  in the vacuum in the Poincar\'e $x^+$ coordinate for the right-moving  modes. The vacuum for the left-moving modes is given by eq.\eqref{vac}. With this choice of vacuum state, and labelling the intersection point of the horizon and the spatial slice $\mathcal{S}$ as $(x_0,x^-)$, we find that $\rho_h$ in eq.\eqref{SentaN} becomes 
 \begin{equation}
 \rho_h=\pqty{\omega-\frac{1}{2}\ln {x^-_v{}'}}\bigg\vert_{h}\label{rhohdef},
 \end{equation}
 where prime denotes a derivative with $x^{-}$ and the horizon is located at $x^+=x_0$.
 Also, from fig \ref{eefig}, the interval \eqref{valhelxp} is given by
 \begin{equation}
 \Delta x^+_v=x_0 - x^-_v(x^-)\label{deint}.
 \end{equation}
 Therefore, the entanglement entropy, \eqref{SentaN}, becomes
 \begin{equation}
 S_{EE}=\frac{N}{6}\pqty{-\ln(\frac{x^+ - x^-}{2})-\frac{1}{2}\ln {x^-_v{}'}+\ln(x^+ - x^-_v)}\bigg\vert_{h}\label{seesec}.
 \end{equation}
Consider the first derivative of $S^{\psi}_{gen}$ along the event horizon, $x^+=x_0$, given by
\begin{align}
     \frac{\text{d}S^{\psi}_{gen}}{\text{d}\lambda}&=(x_0 - x^-)^2\frac{\partial S^{\psi}_{gen}}{\partial x^-}=(x_0 - x^-)^2\pqty{\frac{\partial S_{BH}}{\partial x^-}+\frac{\partial S_{EE}}{\partial x^-}}\label{stotder}.
\end{align}
 Using the forms of $S_{BH}$, eq.\eqref{defentbh},\eqref{gendilq}, and  $S_{EE}$, eq.\eqref{seesec}, we get
 \begin{equation}
 \frac{\text{d}S^{\psi}_{gen}}{\text{d}\lambda}=\frac{(x_0 - x^-)^2}{4G}\pqty{ \frac{h}{(x_0-x^-)^2}+\frac{h'}{x_0-x^-}+\frac{h''}{2}+\frac{\zeta}{2}\pqty{-\frac{x^{-}_v{}''}{2 x^{-}_v{}'}-\frac{x^{-}_v{}'}{x_0-x^{-}_v}+\frac{1}{x_0 - x^-}} }\label{del1sps}.
 \end{equation}
 Using $x^-_v$ given in eq.\eqref{vac} and the late time form of $h$, eq.\eqref{hformp}, we see that at late times,
 \begin{align}
      \frac{\text{d}S^{\psi}_{gen}}{\text{d}\lambda}\rightarrow 0\label{delSpsi}.
 \end{align}
Consider the second derivative of the entropy along the horizon given by 
 \begin{align}
     \frac{\text{d}^2 S^{\psi}_{gen}}{\text{d}\lambda^2}&= (x_0 - x^-)^2 \partial_- \pqty{(x_0 - x^-)^2 \partial_- S^{\psi}_{gen}}\label{2derS}.
 \end{align}
Using  eqs.\eqref{gendilq} and \eqref{seesec}, we get
\begin{align}
\frac{d^2 S_{gen}^{\psi}}{\text{d}\lambda^2}&=\frac{(x_0 - x^-)^4}{8G}\pqty{h'''+\frac{2\zeta}{x_0 - x^- }\pqty{\frac{x^-_v{}''}{2 x^-_v{}'}+\frac{x^-_v{}'}{x_0 - x^-_v}}}\nonumber\\
+&\frac{(x_0 - x^-)^4}{8G}\zeta\pqty{\frac{1}{2}\pqty{ \frac{x^-_v{}''}{x^-_v{}'}}^2-\frac{x^-_v{}'''}{2 x^-_v{}'}-\frac{(x^-_v{}')^2}{\left(x_0 - x^-_v\right){}^2}-\frac{x^-_v{}''}{x_0-x^-_v}-\frac{1}{\left(x_0-x^-\right)^2}}.\label{sgen2der}
\end{align}	
We can simplify the above expression by using the equation of motion for $h$, eq.\eqref{her}, which using $T^{\psi}_{--}$ in the form eq.\eqref{Tpsi}, can be written as
\begin{equation}
h'''=\frac{\zeta}{2}\pqty{\frac{x^-_v{}'''}{x_v^{-'}}-\frac{3}{2}\pqty{\frac{x^-_v{}''}{x^-_v{}'}}^2}-16	\pi G T^m_{--}\label{heqsch}.
\end{equation}

Therefore, using eq.\eqref{heqsch}, the expression eq.\eqref{sgen2der} simplifies to
\begin{align}
     \frac{\text{d}^2 S^{\psi}_{gen}}{\text{d}\lambda^2}&=-\frac{(x_0 - x^-)^4}{8G}\pqty{ \frac{64 G^2 }{\zeta} \pqty{\del_-S_{EE}}^2+16\pi G\, T_{--}^m}\label{del2Spsi},
\end{align}
which is manifestly negative if the energy condition eq.\eqref{energycond} is satisfied. 
From  eq.\eqref{delSpsi} it then follows that $S_{gen}^\psi$ satisfies the condition
\be
\label{genseca}
{\text{d}S_{gen}^\psi\over \text{d}\lambda} >0,
\ee
along the event horizon and asymptotically vanishes as $\lambda \rightarrow \infty$. This shows 
that the generalised entropy satisfies the second law along the future event horizon of the black hole.

\section{Generalised Entropy and Q-Screens  }
\label{qscreens}
We saw in subsection \ref{clasappar} that in the classical theory the area of the apparent horizon also increases monotonically. More precisely, consider the co-dimension one  surface which is foliated by  marginally trapped surfaces (called the future Holographic screen in \cite{boussonewarea}). Then it was shown in subsection \ref{clasappar} that this surface is space-like and the  area of the marginally trapped surface (i.e. the apparent horizon)  increases monotonically as one goes outward, towards the boundary, along it \footnote{As mentioned in subsection \ref{vacuumsols} the horizon value of the dilaton plays the role of the area in the JT model.}.
We also commented towards the end of subsection \ref{chisecondlaw} that the area of  the apparent horizon in the $\chi$ system did not grow in such a monotonic manner. 

Here we consider the behaviour of the generalised entropy, instead of the area of the apparent horizon. A `` future Q-screen" can be defined which is  the analogue of the future holographic screen  with the generalised entropy playing the role of the classical area. 
The idea is as follows, \cite{bousso1506qfc}. In the 2D spacetime we are considering here, we first define the quantum expansion to be the rate of change of the generalised  entropy along a null ray. The   marginally quantum  trapped surface is then defined as  a point on a Cauchy surface at which, a)  the quantum expansion along the outward directed future null ray vanishes, and b) the quantum expansion along the inward  directed future null ray is negative \footnote{In the 2D case being considered here the marginally quantum trapped surface is  a point in spacetime. }. 
With our choice of coordinates, this means 
\begin{align}
\partial_- S_{gen} & =  0 \label{aunt}, \\
\partial_+S_{gen} & <  0 \label{auntplus}.
\end{align}
Finally, the future Q-screen is defined as a surface foliated by  marginally quantum trapped surfaces. 

We show below that the generalised entropy is a monotonic function along a future Q screen. 
This result follows in a very straightforward manner in 2D from another relation, called  the Quantum Focussing  Condition (QFC) which  has also been discussed in the literature, \cite{bousso1506qfc}. 

In 2D the QFC takes the form, 
\be
\label{qfca}
{\text{d}^2S_{gen}\over \text{d}\lambda^2}<0,
\ee
where $\lambda$ is an affine parameter along a null geodesic. 
A related condition is called the Quantum Null Energy Condition (QNEC) which  is given by 
\be
\label{qneca}
{\hbar \over 2 \pi} {\text{d}^2 S_{EE} \over \text{d}\lambda^2}   \le  \langle T_{ab}\rangle k^a k^b  ,
\ee 
with $\langle T_{ab}\rangle$  being  the stress tensor which appears on the RHS of Einstein's equations, and $k^a$ is the tangent vector along a null geodesic, eq.\eqref{tannull} .

We will show below that eq.(\ref{qfca})  is true in both the $\chi$ and $\psi$ systems, when the additional classical matter we add satisfies the NEC eq.\eqref{neca} .  

We also show that the QNEC eq.(\ref{qneca})  holds in both systems when the classical matter  satisfies this condition. 
This will also allow us to establish that the generalised entropy is monotonic along a future Q-screen in both systems. 


\subsection{The $\boldmath{\chi}$ System}
We start with the $\chi$ system and first show that QFC is true. 
We take the definition of the generalised entropy  in this case to be eq.(\ref{entropchi})

Next, we take  an outward directed  null geodesic  along increasing values of $x^-$. 
It  then follows from eq.(\ref{entropchi}), and the discussion in subsection  \ref{chisecondlaw}, see eq.(\ref{mme}) and eq.(\ref{final}), that 
the QFC  eq.(\ref{qfca})  follows for matter satisfying the NEC, eq.\eqref{neca}. The analysis for a null geodesic along the $x^+$ direction is entirely analogous, as long as the condition $T^m_{++}\ge 0$ is also met. 

The role of the entanglement entropy in the $\chi$ system is played by 
\be
\label{entrole}
S_{EE}\rightarrow -{N\over 6} \chi.
\ee
Therefore QNEC eq.\eqref{qneca} takes the form (with $\hbar$ set equal  to $1$):
\be
\label{qnecchi}
{-N\over 12 \pi} {\text{d}^2 \chi \over \text{d}\lambda^2} \le T_{ab} k^a k^b.
\ee
From  eq.\eqref{entropchi}  it follows that QFC eq.(\ref{qfca}) implies
\be
\label{qnecab}
{N\over 12 \pi} {\text{d}^2 \chi \over \text{d}\lambda^2} \ge {1\over 8 \pi G} {\text{d}^2 \phi \over \text{d}\lambda^2}.
\ee
From the equations of motion for $\phi$ it also follows that  
\be
\label{oomph}
{\text{d}^2\phi \over \text{d}\lambda^2}= -8 \pi G T_{ab} k^a k^b.
\ee
Note here $T_{ab}$ is the full stress tensor including the contribution from the $\chi$ field. 
Combining eq.(\ref{qnecab}) and eq.(\ref{oomph}) gives the QNEC condition eq.(\ref{qnecchi}).
We have set $\hbar=1$ above, but it is easily restored as follows. The prefactor $N$ 
in the $\chi$ action, eq.(\ref{nonimpact}) arises from the conformal anomaly for the matter fields and should therefore actually be $N \hbar$, this gives exact agreement with the QNEC \footnote{We also note that our normalisation for the stress tensor is correct; as a check   this is the normalisation which arises after dimensional reduction from $4$ dimensions.}.


We now turn to showing that $S^{\chi}_{gen}$ increases monotonically along a future Q-screen. 
From the QFC it follows that $\nabla_-\nabla_- S^{\chi}_{gen} \le 0$.
Using eq.\eqref{final} and equation of motion in the form eq.(\ref{mme}), it can be seen that along a future Q-screen, where  eq.(\ref{aunt}) is met, 
\be
\label{condos}
\partial_-^2S^{\chi}_{gen} =-\frac{1}{4G}\pqty{\frac{\zeta}{2}(\partial_{-}\chi)^{2}+8\pi G \,T^m_{--}} <0.
\ee
It then follows that no two points on a future Q-screen can have the same value of $x^+$. For if they did, the two points could be connected by a null ray along the $x^-$ direction, and since eq.(\ref{condos}) is true, $\partial_-S^{\chi}_{gen}$ could not  vanish at both points. 
We can therefore use the coordinate $x^+$ to parametrise  points on the future Q-screen. 
The monotonicity of $S^{\chi}_{gen}$ along the Q-screen then follows simply by noting that eq.(\ref{auntplus})  is satisfied on it. 

In fact we can say   more about  future Q-screens  in the $\chi$ system.

For the system in  initial Poincar\'e vacuum state and infalling   null matter of the kind considered in subsection \ref{infallchi} the general solution is given in eq.\eqref{gendilq}, eq.\eqref{valchife}. 
It then follows that 
\be
\label{plus}
\partial_+S^{\chi}_{gen}= -\frac{1}{4G}\pqty{\frac{h}{(x^+ - x^-)^2}+\frac{\zeta}{2(x^+ - x^-)}} <0,
\ee
(since $h>0$, as discussed in appendix \ref{latehap} ). 

This shows that $S^{\chi}_{gen}$ increases monotonically as one goes along the Q-screen towards the boundary. 

In fact it  can be easily seen that the future Q-screen in this case is a space-like surface. This follows by  noting  that the `+-' equation of motion eq.\eqref{pm}, where $T_{+-}^m$ is replaced by eq.\eqref{Tpmchi}, imply that 
\begin{equation}
\partial_+\partial_-S^{\chi}_{gen}=-\frac{1}{8G}\,e^{2\omega}\phi<0\label{qcdpms}.
\end{equation}
Since eq.\eqref{condos} is also true it then follows from eq.\eqref{aunt}  that, along the Q-screen,
\be
\label{nta}
{\text{d}x^- \over \text{d}x^+}= -{\partial_+\partial_- S^{\chi}_{gen} \over \partial_-^2 S^{\chi}_{gen}}<0,
\ee
thereby showing  that the future Q-screen   is space-like.

\subsection{The ${\psi}$ System}
\label{psiqscreen}
The generalised entropy which we denote by $S_{gen}^\psi$ in this case is defined in eq.(\ref{genet}). It follows from discussion in subsection \ref{psisecondlaw}, see eq.\eqref{del2Spsi},  that this generalised entropy also satisfies the condition 
\be
\label{qfcpsi}
{\text{d}^2 S^{\psi}_{gen}\over \text{d}\lambda^2} <0,
\ee
for null geodesics along the $x^-$ direction and classical matter satisfying the energy condition eq.\eqref{energycond}. Similarly for null geodesics along the $x^+$ direction, one can also show that eq.(\ref{qfcpsi}) is true as long as the condition $T^m_{++}\ge0$ is met. 

It also follows in a straightforward way from eq.\eqref{qfcpsi} and the equations of motion for the dilaton that the QNEC is valid in this case. 
The reasoning is entirely analogous to steps eq.\eqref{qnecchi},\eqref{qnecab},\eqref{oomph} in the $\chi $ system above. 

We now turn to the behaviour of generalized entropy along a future Q-screen in this system. 
Since eq.(\ref{aunt}) is met on a future Q-screen, it follows from eq.(\ref{qfcpsi}) that 
$\partial_-^2S^{\psi}_{gen}<0$. It then follows, in a manner analogous to the discussion in the $\chi$ system that the Q-screen can be parametrised by the $x^+$ coordinate and 
that $S^{\psi}_{gen}$ is monotonically varying along the Q-screen.   

It is worth mentioning that, in contrast with the $\chi$ system, the condition  $\partial_+S^{\psi}_{gen}<0$ is not obviously met along a quantum marginal surface where 
$\partial_-S^{\psi}_{gen}=0$. 

We end with one final comment. One can also examine the existence of quantum extremal surfaces \cite{Engelhardt:2014gca}, where both conditions, 
\be
\label{quaintest}
\partial_-S_{gen}=\partial_+ S_{gen}=0,
\ee
are met. 
In the   Kruskal extension of the Schwarzschild geometry (with two boundaries) one finds that there are additional spacetime points where 
eq.(\ref{quaintest}) is satisfied. However the value of $S_{gen}$ at these points is bigger than that at the bifurcate horizon. 
Thus the minimum value for $S_{gen}$ is obtained at the bifurcate horizon and equals the generalised entropy for the black hole, eq.\eqref{entbchi}, eq.\eqref{genenta}, as one would expect on physical grounds.

\section{Conclusions}
\label{conclusions}
In this paper we have considered the JT model coupled to matter in the semiclassical approximation. 
Two different models were analysed, in one case involving a  matter field,  $\chi$, with a non-minimal coupling, and in the other  case  $N$ scalar fields, $\psi_i$,  $i=1, \cdots N$, which are massless and minimally coupled. While in  both cases the matter has the same conformal anomaly, the boundary conditions which are imposed result in differences in the full matter stress tensor  and   the resulting behaviour of the two systems is also then different, see end of subsection \ref{psilaws}. 

In the $\psi$ case the effects of  matter, in the semiclassical limit, is to renormalise the coefficient of the Schwarzian action as discussed in section \ref{psi} \footnote{More correctly this was shown to be true at small temperature ${T \over \mathcal{J}} \ll \phi_B$, and small frequency, ${\omega \over \mathcal{J}} \ll \phi_B$ in section \ref{psi}.}. 
As a result the thermodynamics and response to additional infalling matter qualitatively stay the same as in the classical case.
In contrast, for the $\chi$ system, analysed originally in \cite{alm}, the effect of matter cannot be understood in this simple manner, and the thermodynamics
and also dynamics change more appreciably. 

Starting with the initial state, which corresponds to the Poincar\'e vacuum,  when additional matter is thrown in, we find that in both cases  a black hole forms, and the system thermalises. In the $\chi$ case, a   quasi-normal mode characterises the 
ring-down to the final black hole geometry, once the matter stops falling in, while in the $\psi$ case the system instantly thermalises.

An interesting feature is that in both cases  the second law of thermodynamics is obeyed in the following non-trivial way. 
One can define a generalised entropy which includes  the area of the horizon, given in the JT theory by the horizon value of the dilaton, and a contribution due to the entanglement of matter. In the $\chi$ system this latter term is given by the horizon value of the $\chi$ field, while in the $\psi$ system it is the entanglement due to the $N$ free scalars across the black hole horizon. We show that this generalised entropy increases monotonically along the future event horizon of the black hole, if the extra infalling matter meets the null energy condition. 

 In earlier work, \cite{boussogencosmo}, a future Q-screen was defined which  is the analogue of the locus of an apparent horizon with the   generalised entropy now playing the role of the horizon area. Interestingly, we find that in both  systems    the Quantum Focussing Condition is satisfied, and as a result the generalised entropy also increases along a future Q-screen. 

We have not been able to analyse general time dependent situation here and have restricted ourselves to cases where  eq.\eqref{condfeq} is met i.e., the 
additional matter falls in slowly. It would be interesting to study more general situations as well. 

We have worked here in the semi-classical limit where the 2D Newton constant, $G\rightarrow 0$,  and $N\rightarrow \infty$, keeping $G N$ fixed. Going forward it would be interesting to also incorporate  the quantum effects  of the gravity-dilaton sector systematically, order by order in $G$, and to compare the results with those obtained in \cite{Stanford:2017thb,  Saad:2019lba}. 

In this context it is  worth keeping the following observations in mind.  
The JT model can be obtained by dimensional reduction from higher dimensions, as discussed in \cite{nayak,Moitra:2018jqs,Moitra:2019bub}. The near horizon  $AdS_2$ geometry  obtained in this way is then glued in to the higher dimensional asymptotically flat or $AdS$ space. It would be worth investigating how  this gluing of the geometry  to the higher dimensional asymptotic spacetime affects the results for the higher loop corrections to the partition function etc. 
Also, on carrying out a dimensional reduction 
 extra terms arise in the action; in particular a term   quadratic in the dilaton of the form, \cite{nayak},
\be
\label{exited}
\Delta S = C_1 \int \text{d}^2 x \sqrt{-g}\, \phi^2.
\ee
Due to these extra terms, the geometry  in the near-horizon region is no longer $AdS_2$ and  departs from it, at the same order as $\phi$, in the parameter $1/\cal{J}$, eq.\eqref{linvardil}.
It was already noted in earlier work that these departures can have  significant effects in the semi-classical limit discussed here \cite{Trivedi:1992vh} . 
Incorporating them  while studying the loop corrections in $G$ could also be important. 

It is worth digressing briefly to remind the reader why  the corrections to the action which arise  can lead to significant effects in the semi-classical limit. 
In general the $2$D metric for a static black hole can be written as 
\be
\label{metexta}
ds^2=f(r) (-dt_s^2+dr_*^2),
\ee
where 
\be
\label{rst}
r_*=-\int {dr \over f(r)}.
\ee
At extremality, $f(r)$ has a second order zero at the horizon $r=r_h$. 
If one calculates the quantum back reaction due to a massless scalar field which is in the vacuum with respect to 
$x^\pm_s=t_s\pm r_*$ one finds that, at the future event horizon, this is given  by 
\be
\label{feh}
T_{\mu\nu} U^\mu U^\nu\sim {f'''\over f'},
\ee
where $U^\mu$ is the $4$-velocity of a freely   infalling observer. 
In the JT model where the geometry is exactly $AdS_2$, $f(r)=({r-r_h\over r_h})^2$ and thus $f'''$ vanishes, so that 
 the RHS vanishes. 
However, more generally once, say,  the effects of the $\phi^2$ term are incorporated, the metric will not be  $AdS_2$ and in general $f'''$ will not vanish as one approaches  the horizon. 
For example for  the $4$D extremal RN case we have 
\be
\label{valf4d}
f(r)={(r-r_h)^2\over r^2},
\ee
and $f'''(r_h)$ does not vanish.
It then follows from eq.(\ref{feh}) that since $f'$ will continue to vanish at the horizon,  due to the second order zero at extremality,  the RHS of eq.(\ref{feh}) will diverge. This means a freely infalling observer will see a diverging energy density at the future horizon. This   signals the possibility that the back reaction due to quantum effects 
will not be small. In fact it was shown in \cite{Trivedi:1992vh}  that  the quantum effects do significantly change the behaviour of the geometry near the horizon in the case of the $2$D system obtained after dimensionally reducing the RN black hole from $4$D.

Returning to the main thread of our  discussion   it would clearly then be worth going beyond the semiclassical limit studied here and incorporating the 
quantum corrections order by order in $G$, keeping some of the observations above also in mind.

In conclusion, it is remarkable that  the JT model, which is  a simple model of gravity,   is  proving to be such a  rich laboratory for studying various aspects of quantum gravity. Recent developments \cite{Saad:2019lba,   Penington:2019npb, Almheiri:2019psf, Stanford:2019vob, Almheiri:20192},   suggest that it can provide interesting insights both to  the information puzzle and towards understanding the Euclidean path integral of quantum gravity. In addition, it could be an important laboratory for understanding the different types of bulk entropy associated with different types of horizons and surfaces. We look forward to these developments with considerable anticipation.

%
%
%
%
%
%
%


\acknowledgments
We would like to thank  Ahmed Almheiri, Matthew Headrick, Gautam Mandal, Juan Maldacena, Shiraz Minwalla,  Aron Wall for valuable discussions.
We thank the DAE, Government of India, for support.  
SPT also acknowledges support from the J. C. Bose fellowship of the DST, Government of India and the organisers of the workshop on  {\it Quantum Information and String Theory 2019}, YITP, June 2019. 
UM thanks the organisers of {\it The 36th Advanced School in Physics on Recent Progress in Quantum Field/String Theory} at the Israel Institute for Advanced Studies, The Hebrew University of Jerusalem, Israel, the {\it Spring School on Superstring Theory and Related Topics} at the Abdus Salam International Centre for Theoretical Physics, Italy,  and the {\it It from Qubit School / Workshop (Quantum Information and String Theory 2019)}  at the Yukawa Institute for Theoretical Physics, Kyoto University, Japan for their warm hospitality while this work was in progress. VV would like to thank the organisers of the Spring School on Superstring Theory and related topics at ICTP, Trieste, for their hospitality during the course of this work.
We gratefully acknowledge support from the Infosys Endowment for Research on the Quantum Structure of Spacetime. 
Most of all, we thank the people of India for generously supporting research in String Theory. 

\appendix

\section{Coordinate Transformations}
\label{appcoord}
In this appendix,  for completeness, we give details of the transformations between the Poincar\'e, Schwarzschild and Kruskal coordinate systems and also discuss the transformation properties of the matter stress tensor.

The Poincar\'e metric in light cone coordinates is given by (\ref{poinmet}),
\begin{align}
\text{d}s^2&=-\frac{4\,\text{d}x^+\text{d}x^-}{(x^+-x^-)^2}\label{poin}.
\end{align}
Consider the vacuum dilaton solution given in eq.\eqref{solid}, with the condition eq.(\ref{muabc}) being met. We can do an appropriate $SL(2,R)$ transformation to get the dilaton solution eq.\eqref{solid} to the form
\begin{align}
\phi={1\over \mathcal{J}}\pqty{\frac{1-\mu\, x^-x^+}{x^+-x^-}}\label{dil1},
\end{align}
The relation between these Poincar\'e coordinates and Schwarzschild coordinates is given by,
\be
x^\pm =\frac{1}{\sqrt{\mu}}\,\text{tanh}({\sqrt{\mu}\,{x}_s^\pm\over 2})\label{rintrans},
\ee
under which the Poincar\'e metric eq.\eqref{poin} becomes
\be
\text{d}s^2=-\frac{\mu\,\text{d}{x}_s^+\text{d}{x}_s^-}{\text{sinh}^2\pqty{{\sqrt{\mu}\over 2}\,({x}_s^+-{x}_s^-)}}\label{bhmet},
\ee
and the dilaton, eq.\eqref{dil1}, transforms to
\be
\phi={\sqrt{\mu}\over \mathcal{J}}\,\text{coth}\pqty{{\sqrt{\mu}\over  2}({x}_s^+-{x}_s^-)}\label{dil2}.
\ee
We can convert the metric eq.\eqref{bhmet} into static coordinates by writing it in terms of the coordinates $t_s$ and $r_{\ast}$, where $x_s^{\pm}=t_s\pm r_{\ast}$. Doing so gives
\be
\text{d}s^2=\frac{\mu}{\text{sinh}^2\sqrt{\mu}r_{\ast}}\,\pqty{-\text{d}t_s^2+\text{d}r_{\ast}^2}\label{bhmet2},
\ee
Using the relation  eq.(\ref{dears}) between $r$ and $r_*$, the metric and the dilaton expressed in the coordinates $t_s,r$ become
\begin{align}
\text{d}s^2&=-(r^2-\mu)\,\text{d}t_s^2\,+\,\frac{\text{d}r^2}{r^2-\mu}\label{bhmet3},\\
\phi&=\frac{r}{\mathcal{J}}.\label{dilr}
\end{align}
We define the Kruskal coordinates $X_K^+,X_K^-$ to be
\begin{align}
X_K^-=e^{\sqrt{\mu}\,x_s^-},\quad X_K^+=-e^{-\sqrt{\mu}\,x_s^+}\label{kruskal}.
\end{align}
The Poincar\'e coordinates $x^{\pm}$ and the Kruskal coordinates $X_K^\pm$ are related as 
\begin{align}
X_K^+&=\left(\frac{\sqrt{\mu}x^+ - 1}{\sqrt{\mu}x^++1}\right),\nonumber\\
X_K^-&=\left(\frac{1+\sqrt{\mu}x^-}{1-\sqrt{\mu}x^-}\right)\label{poinkrus}.
\end{align}
Using the transformation eq.\eqref{poinkrus}, the Poincar\'e metric eq.\eqref{poin} becomes
\be
\text{d}s^2 =-\frac{4\,\text{d}X_K^+\text{d}X_K^-}{(1+X_K^+ X_K^-)^2}\label{poinmet2}.
\ee

Next let us discuss how the stress tensor of a massless free scalar field transforms under a simultaneous change of  coordinates and the vacuum. 
We consider two coordinate systems denoted by $(X^+,X^-)$ and $(x^+,x^-)$ with metric,
\begin{align}
\mathrm{d}s^2 & =  - F(X^+,X^-)  \mathrm{d} X^+ \mathrm{d} X^- \label{metrics},\\
\mathrm{d} s^2 & =  -f(x^+,x^-) \mathrm{d}x^+ \mathrm{d}x^- \label{metricsb}.
\end{align}
The stress tensor in the $X^\pm $ coordinates when the system is in the $X^\pm$ vacuum, i.e. annihilated by positive frequency modes of $X^\pm$, is given by \cite{Davies:1976ei, Christensen:1977jc, Birrell:1982ix}
 \begin{align}
\langle X|T_{\mu\nu}|X\rangle & =  \Theta_{\mu\nu}+\frac{1}{48\pi} R\,\,g_{\mu\nu} ,\label{forma} \\
\Theta_{\pm\pm} & =  -\frac{1}{12\pi} \,F^{\half}\,  \del_{X^\pm}^2F^{-\half}, \quad \Theta_{+-}=0.\label{transstress}
\end{align}
Using eq.\eqref{transstress}, let us compute the value of the stress tensor in the coordinates $x^{\pm}$.
Comparing eq.\eqref{metrics} and eq.\eqref{metricsb}, we see that
\begin{align}
f&=F\,\frac{\text{d}X^+}{\text{d}x^+}\,\frac{\text{d}X^-}{\text{d}x^-}\label{mettrans}.
\end{align}
The stress tensor transforms as a rank 2 tensor and so we have
\begin{align}
\langle X | T_{x^+x^+} |X \rangle & =-\frac{1}{12\pi} \pqty{\frac{\text{d}X^+}{\text{d}x^+}}^2\,F^{\half}\,\del_{X^\pm}^2F^{-\half}\nonumber\\
&=-\frac{1}{12\pi}\pqty{\frac{\text{d}X^+}{\text{d}x^+}}^2\,\pqty{\frac{f}{\frac{\text{d}X^+}{\text{d}x^+}\,\frac{\text{d}X^-}{\text{d}x^-}}}^{\half}\,\frac{\text{d}x^+}{\text{d}X^+}\,\frac{\del}{\del x^+}\,\pqty{\frac{\text{d}x^+}{\text{d}X^+}\,\frac{\del}{\del x^+}\,\pqty{\frac{f}{\frac{\text{d}X^+}{\text{d}x^+}\,\frac{\text{d}X^-}{\text{d}x^-}}}^{-\half}}\label{stresstrans}.
\end{align}
Simplifying, we get
\begin{align}
\langle X |T_{x^+x^+}| X\rangle &=-\frac{1}{12\pi}\,f^{\half}\,\del^2_{x^+}f^{-\half}-\frac{1}{24\pi}\pqty{\frac{(X^+)'''}{(X^+)'}-\frac{3}{2}\pqty{\frac{(X^+)''}{(X^+)'}}^2}\nonumber\\
&=-\frac{1}{12\pi}\,f^{\half}\,\del^2_{x^+}f^{-\half}-\frac{1}{24\pi}\,\text{Sch}(X^+,x^+),\label{stresscomp}
\end{align}
where $(X^+)'=\frac{\text{d}X^+}{\text{d}x^+}$. An  analogous relation exists for `$--$' components by replacing all `$+$' indices with `$-$' indices.
 The first term in the final expression, using eq.\eqref{forma},eq.\eqref{transstress}, can be interpreted as value of the stress tensor in the coordinates $x^\pm$ with the system also being in the vacuum with respect to the $x^\pm$ coordinates, that is
 \begin{equation}
 \langle x|T_{x^+x^+}|x\rangle = -\frac{1}{12\pi}f^{\frac{1}{2}}\del^2_{x^+}f^{-\frac{1}{2}}\label{txvac}.
 \end{equation}
 Combining eq.\eqref{stresscomp} and eq.\eqref{txvac}, we get 
 \begin{equation}
 \langle X | T_{x^+ x^+} | X \rangle = \langle x | T_{x^+ x^+} | x \rangle -\frac{1}{24\pi} \text{Sch}(X^+,x^+)\label{strtran}.
 \end{equation}
There is an analogous formula with $(X^+,x^+)$ being replaced by $(X^-,x^-)$.

\section{ADM Mass}
\label{adm}
In this appendix, we give the details of the computation of the ADM mass.
Following the the standard holographic renormalisation methods (see, for instance, \cite{Henningson:1998gx,  Balasubramanian:1999re, Myers:1999psa, Emparan:1999pm, Mann:1999bt, Kraus:1999di, deHaro:2000vlm, Bianchi:2001kw, Skenderis:2002wp}), the ADM mass can be computed from the boundary stress tensor obtained by varying the action with respect to the boundary metric. The boundary is specified by a fixed value of dilaton, eq.\eqref{bdrydil}.
In Poincar\'e coordinates eq.\eqref{poinmet}, this in general will correspond to some trajectory $(t,z(t))$.

To compute the ADM mass, it is convenient to work in the Fefferman-Graham (FG) coordinates which we denote by  $(\hat{t},\hat{z})$.
The metric in these coordinates  takes the form
\be
\label{meteora}
\mathrm{d}s^2 = g_{\hat{t}\hat{t}}\, \mathrm{d} \hat{t}^2+\frac{ \mathrm{d} \hat{z}^2}{\hat{z}^2},
\ee

Asymptotically, as ${\hat z}\rightarrow 0$, the metric and dilation satisfy the conditions,
\begin{align}
g_{\hat{t}\hat{t}} & =   {1\over {\hat z}^2} (1+ \mathcal{O}({\hat z}^2))\label{condor} .\\
\phi & =  {1\over \mathcal{J} {\hat z}} + \mathcal{ O} ({\hat z}^0)\label{phfgb}.
\end{align}

\subsection{Fefferman-Graham coordinate transformation}
\label{fgct}
Let the coordinate transformation from the Poincar\'e coordinates to the FG coordinates be denoted by,
\begin{align}
t&=H(\hat{t})+\hat{z}^{2}G(\hat{t}) + \cdots,\nonumber\\
z&=\hat{z}K(\hat{t})(1+\hat{z}^{2}J(\hat{t}) + \cdots)\label{coordtrans2},
\end{align}
where $H,G,J,K$ are functions which will be determined by imposing FG gauge and also by requiring that the boundary, to leading order, corresponds to  constant $\hat{z}$ .
For ease of notation, the arguments of the functions $G,H,J,K$ will not be written explicitly in the rest of the discussion and it is understood that they are only functions of the time $\hat{t}$. In this appendix the derivatives with respect to $t$ will be denoted by primes and derivatives with respect to $\hat{t}$ will be denoted by dots. Also, we set $L_2=1$.

For the kind of situations we consider, it follows from the equations of motion that the general form of $\phi$  near the boundary 
in Poincar\'e coordinates is given by
\begin{equation}
\phi=\frac{\zeta}{4}+\frac{f_{0}(t)}{z}+f_{2}(t)z+\mathcal{O} (z^2)\label{phiform}.
\end{equation}
(In the classical case discussed in section \ref{basic}, the first term on the RHS is absent).

Expanding in FG coordinates, we get 
\begin{equation}
\phi=\frac{\zeta}{4}+\frac{f_{0}(t)}{K\hat{z}}\Big\vert_{t=H}+\hat{z}\left(\frac{G {f}_{0}'(t)-J f_{0}(t) }{K}+K f_2(t)\right)\Big\vert_{t=H}+\mathcal{O}(\hat{z}^{2})\label{diltrans}.
\end{equation}
Requiring that the dilaton goes like $\frac{1}{\mathcal{J}\,\hat{z}}$ to leading order fixes $K$ to be 
\begin{equation}
K=\mathcal{J}\,f_{0}(t)\label{Kdef}.
\end{equation} 
Note that in eqs. (\ref{diltrans}) and (\ref{Kdef}), the coordinate $t$  in $f_0(t), f_2(t)$ has to be treated as a function of $\hat t$, i.e., $t=H(\hat{t})$. 
The metric under the transformation eq.\eqref{coordtrans2} becomes,
\begin{align}
\text{d}s^2 &=\left(-\frac{\dot{H}^2}{\hat{z}^2 K^2}+\frac{2 J \dot{H}^2+\dot{K}^2 -2 \dot{G} \dot{H}}{K^2}\right)\text{d}\hat{t}^{2}\nonumber\\
&+\left(\frac{1}{\hat{z}^2}+4 J-\frac{4 G^2}{K^2}\right)\text{d}\hat{z}^{2}+\left(\frac{-4 G \dot{H}+2 K \dot{K}}{\hat{z} K^2}\right)\text{d}\hat{t} \, \text{d}\hat{z}\label{transmet}.
\end{align}
Imposing the FG gauge, eqs. (\ref{meteora}), (\ref{condor}), we obtain
\begin{align}
\dot{H}&=K=\mathcal{J}\,f_0(t)\label{Hdef},\\
G&=\frac{K \dot{K}}{2\dot{H}}=\frac{K {K}'}{2}=\mathcal{J}^2\,\frac{f_{0}{f}'_{0}}{2}\label{Gdef},\\
J&=\frac{G^{2}}{K^{2}}=\mathcal{J}^2\,\frac{{f}_{0}'{}^{2}}{4}\label{Jdef}.
\end{align}
Using eqs.(\ref{Kdef}), (\ref{Hdef})-(\ref{Jdef}), the dilaton (\ref{diltrans}) and the metric (\ref{transmet})  become
\begin{align}
\phi&=\frac{\zeta}{4}+\frac{1}{\mathcal{J}\,\hat{z}}+\mathcal{J}\,\hat{z}\left(\frac{{f}_{0}'{}^{2}}{4} + f_{0}f_{2}\right)+\mathcal{O}(\hat{z}^{2})\label{phifg},\\
\text{d}s^{2}&=-\left(\frac{1}{\hat{z}^2}-\mathcal{J}^2 \left(\frac{{f}_{0}'{}^{2}}{2}-f_0\,{f}_0''\right)+\mathcal{O}(\hat{z}^{2})\right)\text{d}\hat{t}^{2}+\left(\frac{1}{\hat{z}^{2}}+\mathcal{O}(\hat{z}^{2})\right)\text{d}\hat{z}^{2}+\mathcal{O}(\hat{z})\text{d}\hat{z}\,\text{d}\hat{t}\label{FGmet}.
\end{align}
It is to be noted that $t=H({\hat t})$ in eq.\eqref{Hdef}-eq.\eqref{FGmet}.  We also note that the coordinate ${\hat t}$ is obtained from $t$  as follows. 
Using eq.(\ref{Hdef}) we can  obtain ${\hat t}$ in terms of $H$ by solving the equation
\be
\label{htrnhatt}
\frac{1}{\mathcal{J}}\int {\mathrm{d}H \over f_0(H)}= {\hat t}.
\ee
From eq.(\ref{Hdef})  we see that this gives   ${\hat t}$ in terms of $t$.

\subsection{Classical Mass}
\label{appclass}
The ADM mass is defined as, 
\begin{align}
M=\lim_{\hat{z}\rightarrow\hat{z}_B}\,\frac{-2 \hat{z}}{\sqrt{-\gamma}}\left(\frac{\delta I}{\delta \gamma^{\hat{t}\hat{t}}}\right)\label{admmass},
\end{align}
where $\hat{z}_B$ is the boundary value of $\hat{z}$ eq.\eqref{bdrydil}, i.e., $\phi_B=\frac{1}{\mathcal{J}\hat{z}_B}+\mathcal{O} (\hat{z}_B^0)$, and $\gamma_{\mu\nu}$ refers to the boundary metric. For the classical case, $I$ is the JT action given in eq.\eqref{jtact}. 
After some calculations, this gives 
\begin{align}
M_{JT}&=\frac{-1}{8\pi G}\lim_{\hat{z}\rightarrow\hat{z}_B}{\hat{z}\gamma_{\hat{t}\hat{t}}}\left(\hat{z}\partial_{\hat{z}}\phi+\phi\right)\label{classm}.
\end{align}
Using the form of the dilaton, eq.\eqref{phifg}, and the metric eq.\eqref{FGmet}, the mass eq.\eqref{classm} becomes 
\begin{equation}
M_{JT}=\frac{\mathcal{J}}{4\pi G}\left(\frac{{f}_{0}'{}^{2}}{4}+f_{0}f_{2}\right)+\mathcal{O} (\hat{z}_B)\label{massjt}.
\end{equation}
Comparing the general vacuum solution for the dilaton, eq.\eqref{solid}, with the form eq.\eqref{phiform} (without the term $\zeta\over 4$ ), we get
\begin{align}
f_0={1\over \mathcal{J}}\,\left(\frac a 2+b\,t+\frac{c t^2}{2}\right),\,\, f_2=-\frac {c} {2\mathcal{J}}\label{fvac}.
\end{align}
Therefore using eq.\eqref{massjt} and eq.\eqref{fvac}, we get mass for the vacuum solution to be eq.\eqref{valmass}.

For a general solution for the dilaton given in eq.\eqref{gendil}, one can derive an expression for the ADM mass in terms of the value of $h$ at the boundary. We expand $h(x^{-})$ in a Taylor series in $z$ near the boundary to obtain,
\begin{equation}
\phi = \frac{h(t)}{2{z}}-\frac{h''(t)}{4}z+\mathcal{O}(z^2)\label{cdile}.
\end{equation}
By comparing  eq.\eqref{phiform} and \eqref{cdile}, we note that 
\begin{equation}
f_0=\frac{h(t)}{2},\quad f_2=-\frac{h''(t)}{4}\label{fval}.
\end{equation}
Expressed in terms of FG coordinates, the dilaton near the boundary takes the form (\ref{phifg}) (without the term $\frac{\zeta}{4}$),
\begin{equation}
\phi=\frac{1}{\mathcal{J}\hat{z}}+\frac{1}{16}\pqty{h'^2-2hh''} \mathcal{J} \hat{z}+ \cdots \label{phfge}.
\end{equation}
From eq.\eqref{massjt} and eq.\eqref{fval}, we obtain the classical mass in terms of $h$ as
\begin{equation}
M=\frac{\mathcal{J}}{64\pi G}\left(h'^{2}-2 h h''\right).\label{admcla}
\end{equation}

From the discussion in section \ref{fgct}, we have seen that asymptotically $t= H({\hat t})$ and therefore from eq.(\ref{Hdef})
and eq.\eqref{coordtrans2},
\be
\label{relthatt}
{\mathrm{d}t \over \mathrm{d} {\hat t}}={\mathcal{J} h(t)\over 2}.
\ee

These observations allow the mass to  be expressed as 
\be
\label{masssch}
M = -\frac{1}{8\pi G \mathcal{J}}\,\text{Sch} (t,\hat{t} ),
\ee
where the Schwarzian derivative on the RHS  is given by eq.\eqref{schderiv}.
We also note that ${\hat t}$ can be taken to be the proper time along the boundary (after suitably rescaling by $z_B$), therefore the Schwarzian
derivative is with respect to the proper time on the boundary.

It follows from eq.(\ref{admmass}) that the mass formula derived above has corrections which are of fractional order 
${\hat z}_B T$ where $T\sim \sqrt{\mu}$
is the temperature. These are small when ${\hat z} _B T \ll 1$. Using eq.(\ref{phfgb}) this condition can be expressed as eq.\eqref{condo}. 
Similarly the mass for the $\chi$ and $\psi$ systems will also be calculated below in the limit eq.\eqref{condo}. 

\subsection{Mass for the $\chi$ system}
\label{amchi}
With the general form of the dilaton given in eq.\eqref{phiform} and the form of the solution for $\chi$, eq.\eqref{solchia}, it is straightforward to see that general form of the dilaton and $\chi$ near the boundary in FG coordinates is given by,
\begin{align}
\phi&=\frac{\alpha_{-1}(\hat{t})}{\hat{z}}+\frac{\zeta}{4}+\alpha_1(\hat{t})\hat{z}+\alpha_2(\hat{t})\hat{z}^2+\cdots\label{phiFG},\\
\chi&=\ln \hat{z}+\sigma_{0}(\hat{t})+\sigma_{1}(\hat{t})\hat{z}+\sigma_{2}(\hat{t})\hat{z}^2+\cdots\label{chiFG},
\end{align}
for some functions $\alpha_i(\hat{t})$ and $\sigma_{i}(\hat{t})$. For the system with $\chi$ field, the action $I$ in eq.\eqref{admmass} is given by
\begin{equation}
I=I_{JT}+I_{\chi}+I_{ct}\label{totact}.
\end{equation}
Here $I_{JT}$ is the JT action given in eq.\eqref{jtact}, and $I_\chi$ is the action for the $\chi$ field defined in eq.\eqref{nonimpact}. $I_{ct}$ is a counterterm which has to be added to cancel the divergences and is given by
\begin{equation}
I_{ct}=-\frac{N}{24\pi}\int_{bdy}\sqrt{-\gamma}\label{sctchi}.
\end{equation}
The mass becomes
\begin{align}
M&=-\lim_{\hat{z}\rightarrow\hat{z}_B}{\hat{z}\gamma_{\hat{t}\hat{t}}}\pqty{\frac{1}{8\pi G}\left(\hat{z}\partial_{\hat{z}}\phi+\phi\right)-\frac{N}{12\pi}\pqty{\hat{z}\partial_{\hat{z}}\chi-\frac{1}{2}}}\nonumber\\
&= \pqty{\frac{\alpha_{1}(\hat{t})}{4\pi G}-\frac{N\sigma_1(\hat{t})}{12\pi}}+\order{(\hat{z}_B)}
\label{genmass}.
\end{align}

Using this we will now derive the formulae for the mass in the cases of eternal black hole and infalling matter. 
For an eternal black hole, the metric in Schwarzschild coordinates is given by eq.\eqref{metic} and the dilaton is given in eq.\eqref{behdil}. In Poincar\'e coordinates, the dilaton becomes eq.\eqref{dil1}. Also, the value of $\chi$ for an eternal black hole is given by eq.\eqref{valchiaa}. 
Converting the dilaton and $\chi$ to FG coordinates and Taylor expanding in powers of $\hat{z}$ near the boundary gives
\begin{align}
\phi&=\frac{1}{\mathcal{J}\,\hat{z}}+\frac{\zeta}{4}+\frac{\mu }{4 \mathcal{J}}\hat{z} + \cdots \label{phiet},\\
\chi&=\ln \hat{z}+\ln( \tilde \phi_B \mathcal{J} )-\sqrt{\mu} \hat{z}+\cdots \label{chiet}.
\end{align}
Therefore, from eqs.\eqref{phiet}, \eqref{chiet}, \eqref{genmass}, we get the mass of the eternal black hole to be eq.\eqref{masschi}.

For the case of an infalling matter pulse in this system , the classical mass eq.\eqref{admclass} gets a
correction due to the contribution from the $\chi$ field. 
We expand the solution eq.\eqref{chisol2} near the boundary in $\hat z$ coordinate,
\begin{align}
\chi&=
\ln{\bigg(\frac{\hat{z}\,\mathcal{J} h}{2}\bigg)}+\ln{\pqty{\frac{2\tilde{\phi}_{B}-h'}{h}}}+\pqty{\frac{h\, h''}{2(2\tilde{\phi}_{B}-h')}+\frac{ h'}{2}}\mathcal{J}\,\hat{z}+\mathcal{O}(\hat{z}^{2})\label{chichie}.
\end{align}
Using eq.\eqref{genmass} and eq.\eqref{chichie}, and taking the limit eq.\eqref{condfeq},  we obtain the mass to be
\begin{equation}
M_{\chi}=\frac{\mathcal{J}}{64\pi G}\left(h'^{2}-2 h h''\right)-\frac{N\, \mathcal{J}\, h'}{24\pi}\label{chiadm2}.
\end{equation}
We can extend the above calculation to the general case where  $\chi$ field  is given by eq.\eqref{valchiabc}. The function $f_-(x^-)$ in eq.\eqref{valchiabc} is determined by imposing Dirichlet boundary condition on $\chi$ at the boundary. Doing so gives $\chi$ as
\begin{equation}
\chi=\ln ({x^+ -x^-\over 2})-\ln(1+a x^+)-\ln({\frac{h(x^-)}{ 2}\over\tilde{\phi}_B-\frac{h'(x^-)}{2}})+\ln(1+a x^- + \frac{a\,h(x^-)}{\tilde{\phi}_B-\frac{h'(x^-)}{2}}).
\label{cgbhd}
\end{equation}
Expanding the above solution for $\chi$ in FG $\hat z$ coordinates  near the boundary we get
\begin{equation}
\chi=\ln{\hat{z} {\cal J} h \over 2}+g_0(\hat{t})+\pqty{\frac{h'}{2}-\frac{a\,h}{2(1+a\,t)}+\frac{(1+at)hh''-2ah(\phi_{B}-\frac{h'}{2})-2a h  h'}{4\pqty{ah+(1+at)\pqty{\tilde{\phi}_B-\frac{h'}{2}}}}}\mathcal{J}\hat{z}+\order{(\hat z^2)}
\label{chbhfg},
\end{equation}
where $g_0{(\hat{t})}$ is a function which is unimportant for the calculation of mass. Using  eq.\eqref{phfge}  (with the additional quantum term, $\zeta\over 4$ as in \eqref{phiform}), eq.\eqref{chbhfg}, the expression for the mass \eqref{genmass} in the limit eq.\eqref{condfeq} becomes
\begin{equation}
M_\chi=\frac{\mathcal{J}}{64\pi G}\left(h'^{2}-2 h h''\right)-\frac{N\, \mathcal{J}\, }{12\pi}\pqty{\frac{h'}{2}-\frac{a\,h}{a\, t+1}}.\label{chiadm3}
\end{equation}
Writing the above expression for the mass in terms of the $t(\hat{t})$ using eq.\eqref{relthatt},  we get
\begin{equation}
M_\chi=-\frac{1}{8\pi G \mathcal{J}}\,\text{Sch}{(t,\hat{t})}-\frac{N\, \, }{12\pi}\pqty{\frac{t''}{t'}-\frac{2 a\,t'}{a\, t+1}}.\label{minbdt}
\end{equation}
The above formula for mass is an $SL(2,R)$ invariant quantity. This can be understood as follows. Under an $SL(2,R)$ transformation,
eq.(\ref{gensl2r}) 
it follows from eq.(\ref{relthatt}) and eq.(\ref{htransf}) (along with  the fact that ${\hat t}$ does not transform under $SL(2,R)$ )  that 
\begin{equation}
t(\hat{t})\rightarrow {p\,t(\hat{t})+q\over r\,t(\hat{t})+s}.\label{tatrans}
\end{equation}
It is then straightforward to verify that the expression for mass eq.\eqref{minbdt} is invariant when $t(\hat{t})$ and $a$ also transform as given in 
eq.(\ref{tatrans}) and eq.(\ref{transformationa}) respectively.

\subsection{Mass for $\psi$ system}
\label{mpsi}
To compute the ADM mass for the $\psi$ system, we once again use the definition eq.\eqref{admmass} with $I$ given by
\begin{align}
I=I_{JT}+I_{\psi}+I_{ct}\label{totact2},
\end{align}
where $I_{JT}$ is the JT action, eq.\eqref{jtact}, $I_{\psi}$ is the action for the matter fields $\psi_i$, eq.\eqref{acts},  and $I_{ct}$ is the counter term action,
which is the same in the $\chi$ system, eq.\eqref{sctchi}. To obtain the quantum correction from $I_{\psi}$, we can examine the conformal anomaly, eq.\eqref{confanom}, for the $N$ scalar fields $\psi_i$ but now including the boundary contribution. It is given by \cite{Polchinski:1998rq,Herzog:2015ioa}
\begin{align}
T^{\mu}_{\mu}=\frac{N}{24\pi}(R+2K\,\delta(x_{\perp}))\label{traceanom},
\end{align}
where $K$ is the extrinsic trace of the boundary. Therefore from the definition of the stress tensor,
\begin{align}
T^\psi_{\mu\nu}=-\frac{2}{\sqrt{-g}}\frac{\delta I_{\psi}}{\delta g^{\mu\nu}}\label{stressdef},
\end{align}
we can obtain the contribution to the mass from $I_\psi$ as follows,
\begin{align}
\lim_{\hat{z}\rightarrow\hat{z}_B}\,\frac{-2 \hat{z}}{\sqrt{-\gamma}}\left(\frac{\delta I_{\psi}}{\delta \gamma^{\hat{t}\hat{t}}}\right)&=\lim_{\hat{z}\rightarrow\hat{z}_B}{\hat{z}}\,T_{{\hat{t}}{\hat{t}}}\big\vert_{bdy}\nonumber\\
&=\frac{N}{12\pi}\,\lim_{\hat{z}\rightarrow\hat{z}_B}\hat{z}\gamma_{\hat{t}\hat{t}}\,K\label{bdytrace}.
\end{align}

Therefore using eq. \eqref{admmass},\eqref{totact2},\eqref{classm},\eqref{phifg} \eqref{bdytrace} and eq.\eqref{sctchi},  the expression for the mass for the $\psi$ system becomes
\begin{align}
M_{\psi}&=-\lim_{\hat{z}\rightarrow\hat{z}_B}{\hat{z}\gamma_{\hat{t}\hat{t}}}\pqty{\frac{1}{8\pi G}\left(\hat{z}\partial_{\hat{z}}\phi+\phi\right)-\frac{N}{12\pi}\pqty{K-\frac{1}{2}}}\nonumber\\
&= \frac{1}{4\pi G}\left(f_{0}f_{2}+\frac{{f}_{0}'{}^{2}}{4}\right)-\frac{N}{12\pi}\lim_{\hat{z}\rightarrow\hat{z}_B}\frac{1}{\hat{z}}\,\pqty{K-1}\label{genmass2}.
\end{align}

From eq.\eqref{FGmet} it can be easily shown that the second term on the RHS above is given by 
\be
\label{st}
\lim_{\hat{z}\rightarrow\hat{z}_B} {1\over {\hat z}} (K-1)= \mathcal{J}^2\,\hat{z}_B\,\left(\frac{{f}_{0}'{}^{2}}{2}-f_0\,{f}_0''\right),
\ee
which involves the same function as appears in the first term in eq.\eqref{genmass2}, when expressed in terms of $h$ using eq.\eqref{fval}.
Thus we see that the quantum effects simply serve to change the coefficient in front of the mass compared to the classical case; this can be incorporated
by rescaling $G$, eq.\eqref{shiftG}. 

It follows then from eq.\eqref{genmass2} that the mass of an eternal black holes is given by 
\begin{equation}
M_\psi=\frac{1}{16\pi G\,\mathcal{J}}\left(1-\frac{\zeta}{2\tilde{\phi}_B}\right)\mu\label{mrespsi},
\end{equation}
 and the mass in the presence of infalling matter is given by 
 \begin{align}
 M_{\psi}=\frac{\mathcal{J}}{64\pi G}\left(1-\frac{\zeta}{2\tilde{\phi}_B}\right)\left(h'^2-2 h h''\right)\label{psiinf}.
 \end{align}
 
From \ref{appclass} we see that the mass can be expressed in terms of the Schwarzian, eq.\eqref{masssch} with $G$ rescaled as given in \eqref{shiftG}. 
%
%

\section{Late time behaviour of $h$}
\label{latehap}
In this appendix, we consider the $\chi$ system  with infalling matter satisfying the condition \eqref{energycond} 
and show that the late time behaviour, after matter has stopped falling in, is given by 
eq.(\ref{valah}). We also show by an explicit coordinate transformation that the late time behaviour eq.\eqref{vale} corresponds to a black hole of the correct mass. 

We first present the arguments for the case of system initially being in the Poincar\'e vacuum. We start at $x^-=0$ with $\phi$ given by   eq.\eqref{valdia}  which corresponds to eq.\eqref{meh} in  eq.(\ref{gendilq}),i.e.,
\begin{equation}
h(0)={1\over \mathcal{J}}, \,\, h'(0)=0,\,\,h''(0)=0\label{hincond},
\end{equation}
and $\chi$ is given by eq.\eqref{achi}
We consider situations where the matter starts falling in at $x^-=0$, and stops falling in at say $x^-=x^-_f$.
We will argue below  that while the matter is falling in  $h$ decreases monotonically and  consider situations where $h$ continues to be  positive
until the instant $x^-_f$. We then argue that for $x^- >x^-_f$, $h$ continues to decrease and eventually hits a first order zero at say $x_0$, in whose vicinity it takes the form, eq.(\ref{valah}). 

To show that while the matter is falling in $h$ decreases monotonically we note the following. 
At $x^-=0$, $h'''<0$, while $h'=h''=0$ from eq.(\ref{hincond}). This means initially, near $x^-=0$, $h'<0$ and $h''<0$. 
Next it can be argued  that the function $h'$ cannot have a minimum. We have noted  that initially $h''<0$,
for it to change sign it would have to go through zero. However we see from eq.\eqref{finaleom} that when $h''=0$, $h'''<0$, thus even if $h''$ were to hit zero,
it  would  subsequently only decrease  and therefore stay negative. Since $h''<0$ it follows that $h'$ stays negative starting from its initially negative value.

Next we will analyse the subsequent behaviour after the matter stops falling in at $x^-_f$.
Before proceeding however, let us make the following observation. If the infalling matter falls in over a  time scale $\tau$ then since $|h|\le {1\over \cal J}$ the condition eq.\eqref{condfeq} is met if 
\be
\label{taucond}
{1\over \tau \cal{J} \phi_B} \ll 1.
\ee
We are taking the infalling matter to obey this condition so that eq.\eqref{condfeq} is valid. 

Once matter stops falling in $h$ satisfies,
\begin{equation}
\label{hint}
h'^2- 2 h h'' -2 \zeta h'=\frac{64\pi G M}{\mathcal{J}},
\end{equation}
 for a constant value of the mass $M>0$, see eq.(\ref{amass}).

To simplify eq.(\ref{hint})  we introduce the variables $\tau$ and $y$ defined by
\begin{align}
\frac{\text{d}}{\text{d}\tau}=h \frac{\text{d}}{\text{d}x^{-}},\quad y=\frac{\text{d}}{\text{d}\tau}\ln{h}=\frac{\text{d}h}{\text{d}x^-}\label{ytaudef}.
\end{align}
It can be shown that  $\tau$ is proportional to the proper time at the boundary, see appendix \ref{adm}.
Rewriting eq.\eqref{hint} in terms of $y$ and $\tau$ using eq.\eqref{ytaudef}, we get
\begin{equation}
\frac{\text{d}y}{\text{d}\tau}-\frac{y^{2}}{2}+\zeta\, y=-\tilde{M},\label{yeq}
\end{equation}
where we have defined 
\be
\label{valtildeM}
\tilde{M}=32\pi G\, {M\over \cal J}.
\ee

The  two roots of the equation
\be
\label{eqq}
y^2 -2\zeta y -2 {\tilde M}=0,
\ee
are 
\begin{align}
y_1& =   \zeta - \sqrt{\zeta^2 + 2 {\tilde M}}\label{valy1}, \\
y_2 & =  \zeta + \sqrt{\zeta^2 + 2 {\tilde M}} \label{valy2},
\end{align}
we see that  $y_1<0$ and $y_2>0$. 
Initially, since we have argued above that $h''<0$, it follows from eq.(\ref{ytaudef})  that right after the matter stops falling in, 
at $x^-=x^-_f$,  ${dy\over d\tau}<0$. 
Thus the value $y$ takes at this time which we denote by $y_f$ satisfies the condition $y_1<y_f<y_2$.
It then follows from eq.(\ref{yeq}) that subsequently $y$ decreases and ultimately reaches $y_1$ with the proper time elapsed being proportional to,

\begin{equation}
\Delta \tau(y) =-\int_y^{y_f} \frac{2}{y^{2}-2\zeta y-2 \tilde{M}}\,\text{d}y\label{tausol}.
\end{equation}
From eqs.\eqref{ytaudef} and \eqref{tausol}, we can obtain $h$ and $x^-$ in terms of $y$ as
\begin{align}
h&=h_0\,e^{U} \quad\text{where}\,\, U=\int \frac{2y}{y^{2}-2\eta y-2\tilde{M}}\text{d}y\label{hyrel},\\
x^{-}&=x_c+\int \frac{2 h_0\,e^{U}}{y^{2}-2\zeta y-2\tilde{M}}\text{d}y\label{xmyrel},
\end{align}
where $h_0$ and $x_{c}$ are integration constants which depend on the  values of $h$, $h'$ and $h''$ at $x^-_f$. 

In the vicinity of $y=y_1$ we get from eq.\eqref{hyrel} that 
\be
\label{behU}
U=\beta_0+\beta_1\ln (y-y_1) + \beta_2(y-y_1) + \cdots,
\ee
where $\beta_0$, $\beta_1$ and $\beta_2$ are constants independent of $y$, 
\begin{equation}
\beta_0= -\frac{2 y_2}{y_1- y_2}\ln{(y_2- y_1)},\,\,\beta_1=\frac{2y_1}{y_1-y_2},\,\,\beta_2=-\frac{2 y_2}{(y_1-y_2)^2}\label{betadef}.
\end{equation}
Using eq.\eqref{behU} in eq.\eqref{hyrel}, eq.\eqref{xmyrel}, we get
\be
\label{behx}
(x_0 - x^-)={2\,h_0\,e^{\beta_0}\over \beta_1(y_2-y_1)} (y-y_1)^{\beta_1}\pqty{1+\frac{\beta_1}{\beta_1+1}(\beta_2-\frac{1}{y_1-y_2})(y-y_1)  },
\ee
\be
\label{behh}
h=h_0\, e^{\beta_0}\, (y-y_1)^{\beta_1}\pqty{1+\beta_2 (y-y_1)+\order{((y-y_1)^2)}},
\ee
where $x^-\rightarrow x_0$ when $y \rightarrow y_1$.
Eq.(\ref{behx}) and (\ref{behh}) give
\be
\label{bexhx}
h=\tilde{c}_1 (x_0-x^-) + \tilde{c}_2 (x_0-x^-)^{2 + \alpha},
\ee
where
\be
\label{valc1t}
\tilde{c}_1=-y_1,
\ee
and 
\be
\label{valaha}
\alpha=-{\zeta\over y_1}={\zeta \over \tilde{c}_1}.
\ee
The form of $h$ in terms of $x^-$,eq.\eqref{bexhx} and  the value of the exponent $\alpha$ in   eq.(\ref{valaha}) agrees with eq.\eqref{valah} and eq.\eqref{valalpha} respectively, when $\tilde{c}_1$ is identified with $c_1\over \mathcal{J}$ in eq.\eqref{valah} .  The value of $\tilde{c}_2$ in eq.\eqref{bexhx} depends on $h_0$ and is unimportant for our discussion.

Let us end this discussion with some comments regarding the case where the system starts from a black hole configuration which corresponds to taking $\chi$ as eq.\eqref{valchiaa} and $\phi$ as in eq.\eqref{valphibha}. The arguments following eq.\eqref{hincond} showing the monotonic decrease of $h$ as matter falls, can be extended to this case in a straightforward manner. 

As matter falls in, $h$ satisfies the equation eq.\eqref{hfbhen}, with the initial conditions 
\begin{equation}
h(0)={1\over \mathcal{J}},\,h'(0)=0,\,h''(0)=-{2\mu\over \mathcal{J}}\label{inicb}.
\end{equation}
From eq.\eqref{hfbhen} and eq.\eqref{inicb} it can be seen that $h'''<0$ at $x^-=0$. Therfore $h''$ continues to decrease starting from the initial value in eq.\eqref{inicb}. It is straightforward to see from eq.\eqref{hfbhen} that when $h''=0$, $h'''<0$ and therefore $h''$ would subsequently decrease as in the case of Poincar\'e vacuum initial condition. Thus $h'$ is always negative and therefore $h$ decreases monotonically.

In the discussion above we have seen that $h$ monotonically decreases. This self consistently justifies neglecting the terms with additional time
derivatives in 
eq.(\ref{tmmchi}).  Each term with an extra derivative  in eq.(\ref{tmmchi}) is also accompanied with an extra power of $h$ on dimensional grounds. 
Since $h={1\over {\cal J}}$ at $x^-=0$, when the matter starts falling in, and decreases subsequently, the condition, eq.(\ref{condfeq})
 suppresses any  term with an extra derivative. Similarly, we see that given eq.(\ref{condo}) the monotonic decrease of $h$ also leads to eq.(\ref{condbs}) 
 being valid, and along with eq.(\ref{condfeq}) then  leads  to eq.(\ref{mmhm}) being  well-approximated by eq.(\ref{hfbhen}).

\subsection{Late time black hole coordinates}
\label{coord}
Here, we show that the late time form of $h$ given in eq.\eqref{vale} corresponds to a  black hole with mass $M$, eq.\eqref{masschi} with $\mu=\frac{c_1^2}{4}$. 

From eq.\eqref{vale} and using eq.\eqref{gendilq}, the dilaton at late times is given by
\begin{equation}
\phi={1\over \mathcal{J}}\pqty{\frac{c_1(x_{0}-x^{-})}{x^{+}-x^{-}}-\frac{1}{2}c_{1}} + \frac{\zeta}{4} \label{philate}.
\end{equation}
By taking 
\be
c_1\rightarrow {2\sqrt{\mu}},\,\,x_0\rightarrow{1\over\sqrt{\mu}},x^\pm\rightarrow {{2}\, x^\pm\over \sqrt{\mu}\, x^{\pm}+{1}}\label{ptop},
\ee
the solution eq.\eqref{philate} can be converted to the standard form for the black hole solution 
\begin{equation}
\phi={1\over \, \mathcal{J}}\pqty{{1-\mu\, x^+ x^-\over x^+ - x^-}} + \frac{\zeta}{4}, \label{phiex}
\end{equation}
which is obtained by taking
\begin{equation}
h={1\over\mathcal{J}}(1-\mu (x^-)^2)\label{etbhh},
\end{equation} in eq.\eqref{gendilq}.  

The coordinate transformation relating Schwarzschild coordinates $t_s,r$ with $x^\pm$ in eq.\eqref{philate} is
\begin{align}
{r}&=\frac{c_1(x_{0}-x^{-})}{x^{+}-x^{-}}-\frac{1}{2}c_{1},\nonumber\\
t_s&=-{1\over c_1}\ln((x_{0}-x^{-})(x_{0}-x^{+}))\label{coordtr}.
\end{align}
This gives for the metric,
\begin{equation}
\text{d}s^{2}=-\frac{4}{(x^{+}-x^{-})^{2}}\text{d}x^{+}\text{d}x^{-}=-\pqty{r^{2}-\frac{c_1^2}{4}}\text{d}{t}_s^{2}+\frac{\text{d}r^{2}}{\pqty{r^{2}-\frac{c_1^2}{4}}}\label{bhlate}.
\end{equation}
Comparing with eq.\eqref{metic} we see that this is a black hole of mass $M$, eq.\eqref{masschi} with the identification eq.\eqref{valc1}. 
The temperature of the black hole eq.\eqref{bhlate} is also easily seen to be
\begin{align}
T=\frac{c_1}{4\pi}\label{tempc1}.
\end{align}

\section{Second law violation for apparent horizon}
\label{cldlt}
In this appendix, we consider the $\chi$ system and show by taking an explicit example for $T_{--}^m$ corresponding to a delta-function pulse that the generalized second law is violated when the generalized entropy is computed along the classical apparent horizon, eq.\eqref{condapph}.

We take the matter stress tensor corresponding to infalling matter to be
\begin{equation}
T_{--}^m=\tilde{\mu} \delta(x^{-})\quad \text{for}\,\,\,\tilde{\mu} >0\label{tmmdelta}.
\end{equation}
 Before the pulse went in, the geometry is pure $AdS_2$, eq.\eqref{poinmet}, and we take the dilaton to be specified by eq.\eqref{valdia}, which corresponds to $h={1\over \mathcal{J}}$. Although the equation for $h$, eq.\eqref{finaleom}, can be solved exactly for the matter stress tensor, eq.\eqref{tmmdelta}, as we saw above in appendix \ref{latehap}, we resort to a perturbative expansion in $\zeta$ for simplicity. The total solution for $h$ can be written as 
 \begin{equation}
 h=h^{(0)}+\zeta h^{(1)}+\order({\zeta^2})\label{hexpan}.
 \end{equation}
  The zeroth order solution  $h^{(0)}$ is obtained by integrating \eqref{finaleom} with the stress tensor eq.\eqref{tmmdelta} and setting $\zeta=0$ which gives
\begin{equation}
h^{(0)}={1\over\mathcal{J}}(1 - \mu(x^-)^2),\label{hosol}
\end{equation}
where 
\be
\label{defvalmb}
\mu=8 \pi G {\tilde \mu \cal{J} }.
\ee

Expanding eq.\eqref{finaleom} to $\order{(\zeta)}$, we get the equation for $h^{(1)}$ as 
\begin{equation}
h^{(1)}{}'''=-\frac{h^{(0)}{}''}{h^{(0)}}\label{firsteom}.
\end{equation}
Using the solution eq.\eqref{hosol}, we can solve for $h^{(1)}$ to get
\begin{equation}
h^{(1)}=-\frac{1}{2\sqrt{\mu}}\pqty{2 \sqrt{\mu}x^{-}+(1-\sqrt{\mu}x^{-})^{2}\ln{(1-\sqrt{\mu}x^{-})}-(1+\sqrt{\mu}x^{-})^{2}\ln{(1+\sqrt{\mu}x^{-})}}\label{persol},
\end{equation}

The perturbative solution is a good approximation everywhere except at late times near  $x^-={1\over\sqrt{\mu}}$ where $h^{(0)}$ goes to zero. 
 The trajectory of the apparent horizon gets corrected due to the correction in $h$. From the general form of the trajectory of the apparent horizon,eq.\eqref{xpxmrela}, and using eq.\eqref{hosol}, eq.\eqref{persol}, we get the apparent horizon trajectory to $\order{(\zeta)}$ to be
\begin{equation}
x^+\vert_{ah}=\frac{1}{\sqrt{\mu}}-{\mathcal{J}\zeta\over 2\mu}\pqty{\sqrt{\mu}x^- - 2\ln(1+\sqrt{\mu}x^-)}\label{qatraj}.
\end{equation}

The dilaton, \eqref{gendilq}, and $\chi$, eq.\eqref{chisol2}  at the apparent horizon to $\order{(\zeta)}$, using the solutions eq.\eqref{hosol}, eq.\eqref{persol}, become 
\begin{align}
\phi\big\vert_{ah}&={\sqrt{\mu }\over \mathcal{J}}+\frac{\zeta}{4}  \left(1-2 \sqrt{\mu } x^-\right)\label{dilappar},\\
\chi\big\vert_{ah}&=-\ln({\sqrt{\mu}\over \mathcal{J}}(1+\sqrt{\mu}x^-))+ O (\zeta)\label{chiapar}.
\end{align}
 Using eq.\eqref{dilappar} and eq.\eqref{chiapar}, the entropy at the apparent horizon is given by 
\begin{align}
S^{\chi}_{gen}&=\pqty{\frac{\phi}{4G}-\frac{N\chi}{6}}\bigg\vert_{ah}\nonumber\\
&=\frac{1}{4G}\pqty{{\sqrt{\mu }\over\mathcal{J}}+\frac{\zeta }{2}\left(\frac{1}{2} - \sqrt{\mu }  x^{-} +\ln({\sqrt{\mu}\over \mathcal{J}}(1+\sqrt{\mu}x^-)) \right)+\order{(\zeta^2)}}\label{entcap}.
\end{align}
%

Ignoring the $\order{(\zeta^2)}$ correction, it is clear from the above expression that entropy is in fact a monotonically decreasing function of $x^-$ and therefore the generalized second law is violated at the apparent horizon. 

The same conclusion can be reached, without taking recourse to perturbation theory in $\zeta$,  by solving eq.\eqref{finaleom} numerically for the case of a delta function matter pulse with initial conditions as given in eq.\eqref{hincond}.

\bibliographystyle{JHEP}
\bibliography{refs}
\end{document}